\documentclass{article}

\usepackage{microtype}
\usepackage{graphicx}
\usepackage{subfigure}
\usepackage{booktabs} % for professional tables

\usepackage{lineno,hyperref}
\usepackage{amsfonts}
\usepackage{amsmath}
\usepackage{indentfirst}
\usepackage{graphicx}
\usepackage{subfigure}
\usepackage{booktabs}
\usepackage{multirow}               %
\usepackage{colortbl}               % 表格颜色的包
\definecolor{mygray}{gray}{.6}
\usepackage{adjustbox}
\usepackage{rotating}               % 横页显示的包
\allowdisplaybreaks[4]

\usepackage{hyperref}

\usepackage[accepted]{icml2019}

\begin{document}

\twocolumn[
\icmltitle{Discovering Differential Features: Adversarial Learning for Information Credibility Evaluation}




\begin{icmlauthorlist}
\icmlauthor{Lianwei Wu}{A}
\icmlauthor{Yuan Rao}{A}
\icmlauthor{Ambreen Nazir}{A}
\icmlauthor{Haolin Jin}{A}
\end{icmlauthorlist}

\icmlaffiliation{A}{Lab of Social Intelligence and Complexity Data Processing School of Software Engineering, Xi'an Jiaotong University,  Xi'an, 710049, China}

\icmlcorrespondingauthor{Yuan Rao}{raoyuan@mail.xjtu.edu.cn}

\icmlkeywords{Machine Learning, ICML}

\vskip 0.3in
]



\printAffiliationsAndNotice{}  % leave blank if no need to mention equal contribution

\begin{abstract}
A series of deep learning approaches extract a large number of credibility features to detect fake news on the Internet. However, these extracted features still suffer from many irrelevant and noisy features that restrict severely the performance of the approaches. In this paper, we propose a novel model based on \textbf{A}dversarial \textbf{N}etworks and inspirited by the \textbf{S}hared-\textbf{P}rivate model (ANSP), which aims at reducing common, irrelevant features from the extracted features for information credibility evaluation. Specifically, ANSP involves two tasks: one is to prevent the binary classification of true and false information for capturing common features relying on adversarial networks guided by reinforcement learning. Another extracts credibility features (henceforth, private features) from multiple types of credibility information and compares with the common features through two strategies, i.e., orthogonality constraints and KL-divergence for making the private features more differential. Experiments first on two six-label LIAR and Weibo datasets demonstrate that ANSP achieves the state-of-the-art performance, boosting the accuracy by 2.1\%, 3.1\%, respectively and then on four-label Twitter16 validate the robustness of the model with 1.8\% performance improvements.
\end{abstract}

% 1. 引言
\section{Introduction}
\label{sec:1}

In recent years, the problem of information credibility has gained much attention and been extremely highlighted. Fake news (a.k.a. hoaxes, rumors, etc.) and misinformation have dominated the news cycle since the US presidential election (2016) \cite{allcott2017social} and even 1\% of users are exposed to 80\% of fake news \cite{Grinberg374}. Meanwhile, in the 2017 German Federal Election, the key factor of AfD party (alternative for Germany party) from no seats to 94 seats in Congress is benefit from flooded automated accounts (a.k.a. bots), which attempts to skew the opinion of real users by inundating them with information that matches the bots' goals \cite{morstatter2018alt}. Furthermore, penetrating research \cite{lazer2018science, vosoughi2018spread} shows that fake news has greater vitality than true information, which diffuses significantly farther, faster, deeper, and more broadly than the truth in all categories of information, and the effects are greatly pronounced. Therefore, the urgency of information credibility evaluation has drawn significant attention in both industries and academia.

% 图1(a-b)
\begin{figure}
\centering
\includegraphics[width=0.4\textwidth]{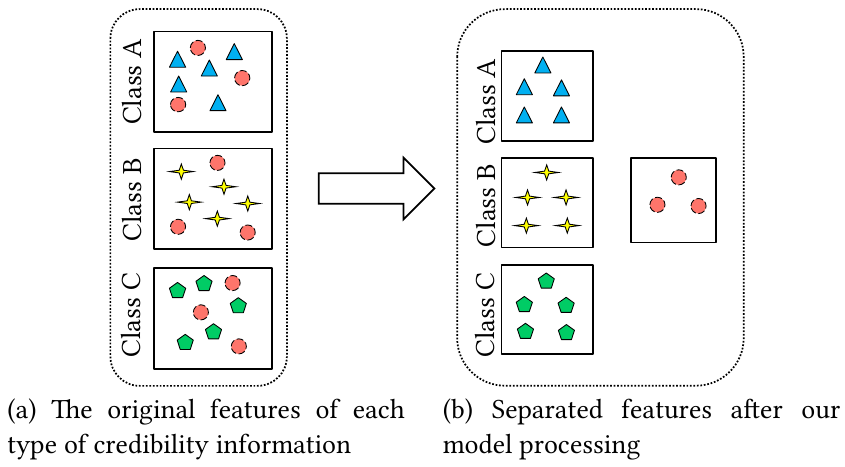}
\caption{The goal of our model for processing features of different types of credibility information. The red circles denote common features. The blue triangles, yellow stars, and green pentagons denote differential features of Class A, B and C, respectively.}
\label{Fig1}
\end{figure}

Most scholars consider credibility evaluation as a text classification problem and devise different deep neural networks to capture credibility features from different perspectives for evaluation have shown to be successful. Specifically, some methods provide in-depth analysis around content features, e.g., linguistic \cite{reis2019supervised}, semantic \cite{de2018attending}, emotional \cite{ajao2019sentiment}, and stylistic \cite{potthast2018stylometric}, and achieve limited performance. On this basis, some work additionally extracts various social context features as credibility features, including meta-data based (i.e., source-based \cite{rathore2017social,yu2018rumor}, user-centered \cite{long2017fake,ribeiro2018Like}, and post-based \cite{wang2017liar, ma2018rumor}) and network-based \cite{ruchansky2017csi, liu2018early, LIU20181}, and promotes the development of different fusion approaches, such as hybrid-CNN model \cite{wang2017liar}, CSI model \cite{ruchansky2017csi}, and tree-structured RNN \cite{ma2018rumor}, which gain remarkable performance boosts compared to other models only capturing text features. From these methods, we can find that expanding features can significantly improve the performance of credibility evaluation.

However, the above methods ignore the fact that many useless, irrelevant, and noisy features will be increased accordingly when the relevant credibility features are gradually added into models, as shown in Figure \ref{Fig1}(a). This not only decreases the performance of models but also leads to a dramatic reduction in the calculation efficiency.

To address the above problems, we propose an \textbf{A}dversarial \textbf{N}etworks method inspired by \textbf{S}hared-\textbf{P}rivate model(henceforth, ANSP) for credibility evaluation, which tries to reduce common and irrelevant-type features from multi-types credibility information. The intuitive description is shown in Figure \ref{Fig1}. In detail, ANSP involves two tasks: Task 1 utilizes adversarial networks guided by reinforcement learning (RL) to extract the important and task-invariant words for capturing common features between true and false information; Task 2, as the target task, first uses BiLSTM to extract credibility features (henceforth, private features) among multiple types of credibility information and then devises a feature separation module including two strategies, i.e., orthogonality constraints and negative KL-divergence, to compare these two types of features for making the private features more differential. Finally, all loss produced from the above steps is merged relying on a linear combination way for optimization.

The main contributions of this paper are summarized as follows:

\begin{itemize}
  \item It is effective for the performance boost of ANSP that we exploit adversarial networks guided by reinforcement learning to sample important, valuable words from input sequences for the capture common features of true and false information, which can improve 4.2\% in accuracy on Weibo compared with all words encoded to roughly and equally share parameters.

  \item We design two strategies to discover differential features, i.e., orthogonality constraints for making common features and private features independent and negative KL-divergence for strengthening the diversity of these types of features. The combination of these two strategies in ANSP achieves 7\% performance improvements on LIAR.

  \item Experiments first on the LIAR \cite{wang2017liar} and Weibo datasets demonstrate that ANSP achieves state-of-the-art performance, outperforming the latest methods by 2\%-3\% in accuracy and then on Twitter16 \cite{ma2017detect} validate the robustness of the model with 1.8\% performance boosts.
\end{itemize}

The remainder of the paper is organized as follows: the next section outlines related work. Section \ref{sec:3} presents the architecture of ANSP and explains the design of each step in details. Experimental results and discussion are described in section \ref{sec:4}, and finally, section \ref{sec:5} summarizes conclusions and future work.

% 2. 相关文献
\section{Related Work}
\label{sec:2}
In this section, we briefly review existing work related to information credibility evaluation and the involved technologies in this paper, including shared-private models, adversarial networks, and reinforcement learning.

\subsection{Information Credibility Evaluation}
\begin{figure}
  \centering
  \includegraphics[width=0.4\textwidth]{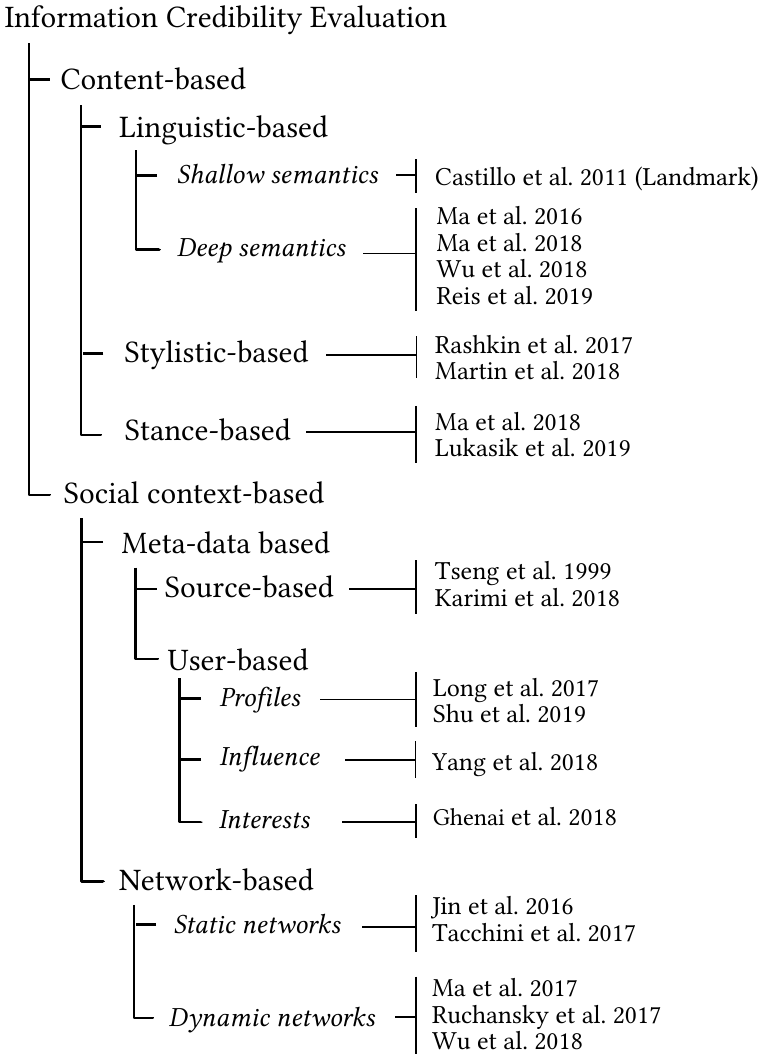}
  \caption{The review of information credibility evaluation methods}\label{fig:2}
\end{figure}

Most of the methods focus mainly on extracting text content features and social context features to evaluate information credibility. The Figure \ref{fig:2} organizes the literature on information credibility evaluation in terms of both types of features.

For content-based methods, features are extracted as linguistic-based, stylistic-based, and stance-based. On linguistic-based methods, a landmark method devised by Castillo et al. \cite{castillo2011information} extracts 68 shallow linguistic features from Twitter to analyze the information credibility of news, e.g., surface text features, shallow semantic features (like the number of keywords, tenses of messages, etc.), emotional features, text topic features, etc. Subsequently, many linguistic-based approaches are developed based on deep learning \cite{reis2019supervised,ma2016detecting,wu2018falseinformation,ma2018rumor,wu2019multi}. Typically, Ma et al. \cite{ma2016detecting} present a deep learning framework for rumor debunking, which learns RNN models by utilizing the variation of aggregated information across different time intervals related to each event. Next, Ma et al. \cite{ma2018rumor} also design two recursive neural models based on bottom-up and top-down tree-structured neural networks to represent rumors, which can learn discriminative features from the propagation structures to detect rumor. Stylistic-based methods usually make great efforts to detect fake news by capturing the manipulators in the writing style of news content \cite{rashkin2017truth,potthast2018stylometric}. As a concrete example, Rashkin et al.\cite{rashkin2017truth} compare the language of real news with that of satire, hoaxes, and propaganda to find linguistic style characteristics of untrustworthy text and present an analytic study on the language of news media in the context of political fact-checking and fake news detection.  Stance-based methods focus on capture stance features as auxiliary features for rumor detection \cite{ma2018detect,lukasik2019gaussian,wu2019different}. For instance, Ma et al. \cite{ma2018detect} propose a multi-task learning framework that unifies rumor detection and stance classification tasks and then extracts the common and tasks-invariant features between both tasks as external features for boosting the performance of both tasks. These methods are effective for specific issues but general issues.

% 图2
For social context-based methods, features are divided into meta-data based (including source-based, user-based, etc.) and network-based. The earlier methods in meta-data based methods for credibility evaluation are source-based \cite{tseng1999credibility,karimi2018multi}, which generally extracts the characteristics of the medium itself to verify credibility based on the principle of ``if the source is reliable, text content is credible". With the popularity of social media, it is difficult to accurately recognize information credibility based on its source credibility. To better evaluate information credibility, more and more credibility features are extracted around social media, especially user features. User-based methods capture disseminator's profile \cite{long2017fake,shu2019role}, influence \cite{yang2018dynamic}, interests features \cite{ghenai2018fake} and adopt various classification models to detect false information on social media, which demonstrate that these features as distinguishing credibility features are useful, and the concatenation of the user features and text features as the input of models achieves average performance improvements margin of about 5\% than only text features as the input of models. Additionally, network-based methods involve static-based and dynamic-based. Static networks are formed by users on social media in terms of interests, topics, and relations, and dynamic networks constructed by propagations of false information also provide many credibility features. Network-based methods \cite{jin2016news,tacchini2017some,ruchansky2017csi,KUDUGUNTA2018312,shu2019Fakenewsnet} generally analyze the networks constructed to study transmission subtree, depth, the degree distribution, clustering coefficient, etc., existing network metrics in order to discover the otherness of true and false information. For example, Jin et al. \cite{jin2016news} exploit conflicting social viewpoints in a credibility propagation network for verifying news automatically in microblogs. By formulating credibility propagation on the network as a graph optimization problem, they define a sensible and effective loss function and provide an iterative optimal solution. In addition, capturing dynamic networks can significantly improve the performance of information credibility. Wu et al. \cite{wu2018tracing} concentrate on modeling the propagation of messages in a social network, which infers embeddings of social media users with social network structures and utilizes an LSTM-RNN model to represent and classifies propagation pathways of a message for identifying fake news and finally improve more than 6\% in macro-F1 and micro-F1.

\subsection{Shared-Private Models}
Recently, the shared-private model is developed by Bousmalis \cite{bousmalis2016domain} for multi-task learning, which strives to separate the input features into common features and private features, where common features in shared space refer to the features that exist simultaneously in different tasks. Conversely, private features represent unique features extracted from different tasks. This model expands the novel idea in many fields, such as spoken language understanding \cite{lan2018semi}, facial detection \cite{trottier2017multi}, and sentiment analysis \cite{buechel2018word}. However, the major limitation of this model is that the shared feature space and the private feature space suffer from feature mixing and redundancy, i.e., some unnecessary task-variant features slip into private space, vice versa. To address this problem, Liu et al. \cite{liu2017adversarial} present an adversarial multi-task learning framework to prevent the shared and private latent feature spaces from interfering with each other and experimental results on 16 different text classification tasks show the benefits of the model. On this basis, another variant is Multinomial Adversarial Networks (MAN) proposed by Chen et al. \cite{chen2018multinomial} for achieving domain adaptation and multi-domain text classification, which is a theoretically sound generalization of traditional adversarial networks that discriminates over two distributions and achieves state-of-the-art performance for domains without labeled data. Even though the improved shared-private models obtain more precise shared and private spaces, the private (task-variant) features extracted from multiple tasks is not sufficiently independent. How to effectively obtain independent task-variant features is one of the valuable questions we solved.

\subsection{Adversarial Networks}
The idea of adversarial networks is initially presented by Goodfellow et al. \cite{goodfellow2014generative} for image generation, besides, and has been applied broadly in many tasks of NLP field, such as information retrieval \cite{kenter2017neural}, machine comprehension \cite{wang2017conditional}, dialog generation \cite{lu2017best}, and fake news detection \cite{wang2018eann}. The goal of adversarial networks is to use a generative network $G$ to generate a data distribution $P_G(x)$ that matches the real data distribution $P_{data}(x)$ as much as possible. Additionally, the model also learns a discriminator $D$ to distinguish $P_G(x)$ and $P_{data}(x)$. Here, it is a min-max game: the model should make the discriminator maximize classification as far as possible, meanwhile, it also makes the generator minimize the gap of the distribution about generative data and real data. The optimized function shows as follows:
\begin{equation}\label{eq1}
\setlength{\abovedisplayskip}{0pt}
\setlength{\belowdisplayskip}{0pt}
\scriptsize
  \epsilon\!=\!min_Gmax_D\!\left\{\!E_{x \sim P_{data}}[ logD(x)]\!+\!E_{z \sim P_{(z)}}[log(1\!-\!D(G(z)))\!] \right\}
\end{equation}
Relying on the advantage of the game of generator and discriminator, our model captures purer common features by preventing Task 1 to classify correctly true and false information.

\subsection{Reinforcement Learning (RL)}

With the fame of AlphaGo, reinforcement learning has become more and more popular in academic communities. The methods of reinforcement learning to solve problems have springing up \cite{lample2017playing,henderson2018deep,conti2018improving,li2019robust}. Due to the embeddings of words in the NLP field are mostly discrete, there is relatively little research to combine reinforcement learning with NLP issues until the SeqGAN \cite{yu2017seqgan} appears. SeqGAN, a sequence generation method, effectively trains generative adversarial nets for structured sequences. Specifically, they model the data generator as a stochastic policy in reinforcement learning and use gradient policy update to solve the generator differentiation problem. The results of the GAN discriminator judged on complete sequence regard as the RL reward, and Monte Carlo search is used to choose the intermediate state-action steps. Subsequently, Zhang et al. \cite{zhang2018learning} also apply reinforcement learning to learn sentence representation by discovering optimized structures automatically. In details, they adopt two representation models: ID-LSTM distills task-words to form purified sentence representation, and HS-LSTM discovers phrase structures to form hierarchical sentence representation. In this work, we rely on the advantages of RL that take trial and error to gain experience for obtaining optimal results, to guide adversarial networks for a job well done.

\section{The ANSP Model}
\label{sec:3}

In this section, we elaborate on the details of ANSP. We first present the overall architecture of ANSP and then explain the specific design of Task 1 via adversarial learning guided by reinforcement learning. Next, we introduce the structure of Task 2 from three parts: private feature extractor, feature separation module, and multiple classifier. Finally, we integrate all loss produced by the above steps into total loss for training. Besides, appendix A lists the main notations and definitions in this paper.

% 图3
\begin{figure*}
  \centering
  \includegraphics[width=0.8\textwidth]{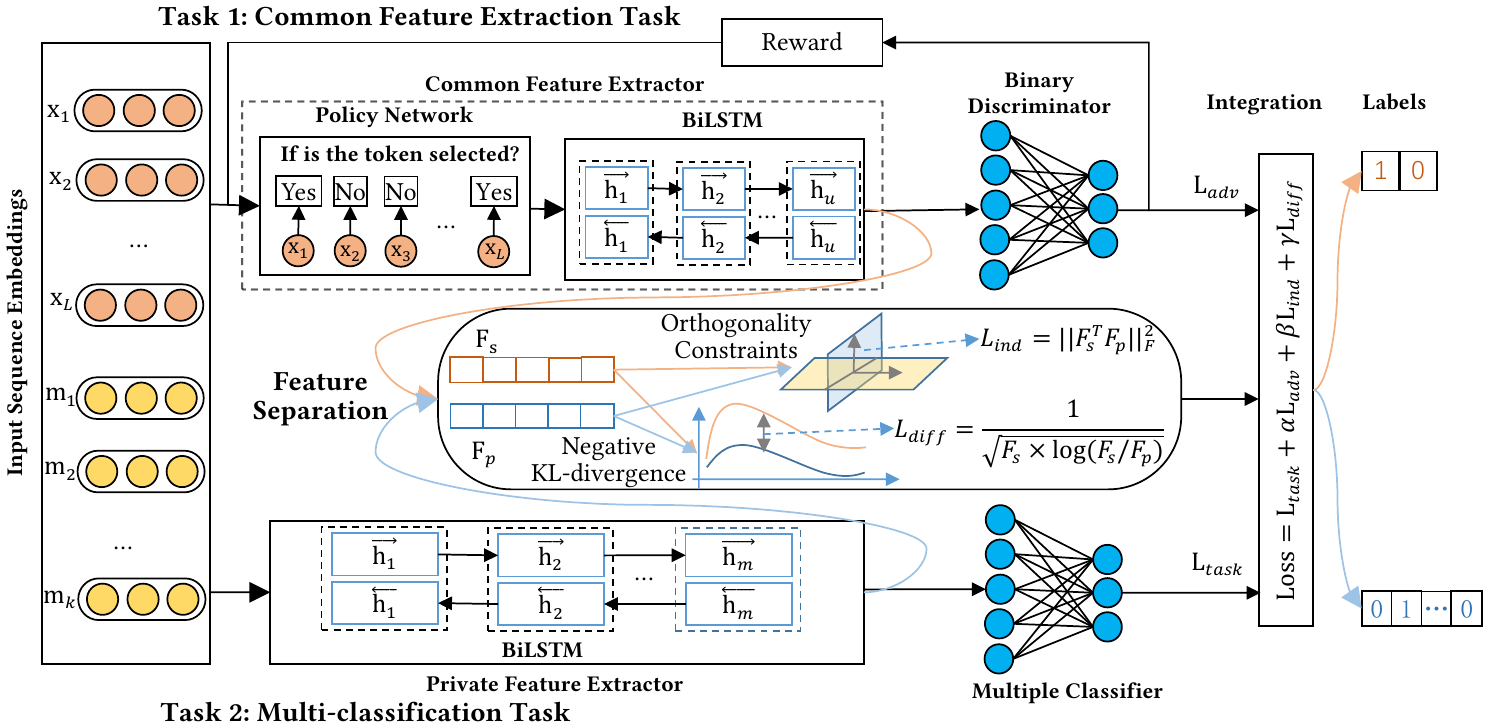}
  \caption{The overview of ANSP. ANSP contains two tasks. Task 1 uses adversarial network guided by reinforcement learning to obtain common features from two types of information: true and false. Task 2 is designed for capturing private features. Feature separation module aims at capturing more differential private features by two strategies to compare both types of features.}\label{fig:3}
\end{figure*}

\subsection{Model Architecture}
The architecture of ANSP is shown in Figure \ref{fig:3}. The goal of ANSP model is to obtain differential private features for information credibility evaluation. \textbf{Task 1} is a binary classification task of true and false information guided by adversarial networks, where common feature extractor acts as the generator to learn features from input sequences for confusing binary discriminator. When a strong binary discriminator cannot classify correctly true and false information, the learned features are essentially common features. To reduce noise caused by insignificant words of input sequences, common feature extractor applies policy network to choose important, valuable words. \textbf{Task 2} obtains private features via BiLSTM for fine-grained information credibility evaluation. To reduce common features among the private features, we design a \textbf{feature separation module} with two strategies to compare the common features from Task 1 with the private features in order to make the credibility features more distinctive. Finally, the model is optimized through the integration of losses generated by the above components.

\subsection{The Inputs of ANSP}
Two tasks of ANSP utilize the same inputs: the concatenations of word embeddings and meta-data embeddings in one tweet $X$, i.e., $X=\{x_1, x_2, ..., x_L, m_1, m_2, ..., m_k\}$. Word embeddings $x_i$ of the specific word $i$ in a tweet text sequence is a $d$-dimensional vector obtained by pre-trained Word2Vec model \cite{Mikolov2013Efficient}. Meta-data embeddings include $k$ types of meta-data (see subsection \ref{sec:4.1} for the details of meta-data). Each type of meta-data $m_i$ is represented to a $k$-dimensional vector by one-hot encoding and then extended a fixed $d$-dimensional vector ($k<<d$) to ensure the same vector length compared with word embeddings. Especially, due to different tweets have different length of word sequences, our practice to addressing variable-length sequences is to take the maximum value $L$ of all sequences in the dataset. Provided that the word sequence of a single tweet is less than $L$, the insufficient part is added zero. Additionally, the concatenation mechanism is ultimately selected as the fusion mechanism of word embeddings and meta-data embeddings, which is simple but effective.

% 3.2
\subsection{Task 1 via Adversarial Learning Guided by Reinforcement Learning}
In Task 1, in order to better capture common features, common feature extractor (including a $\theta$-parameterized chosen policy $G_\theta$) adopts reinforcement learning to guide adversarial networks for choosing the important, invariant words as generating features. Meanwhile, a $\phi$-parameterized binary discriminative model $D_\phi$ is trained to test the performance of $G_\theta$. We interpret the detailed procedure from the perspective of reinforcement learning.

In timestep $t$, the state $s_{t-1}$ encodes the input token $x_{t-1}$ and previous tokens, the action $a_t$ is to choose or ignore the current input token $x_t$, the chosen policy $G_\theta (a_t|s_{t-1};\theta)$ is a stochastic policy. The initial state $s_1$ encodes the first input token $x_1$, where $a_1$ is the random initial action. Our goal is to utilize the chosen policy $G_\theta$ (a.k.a. the generated policy in this paper) to capture important, task-invariant words from input features for extracting common features. In particular, BiLSTM is selected as the encoder of common extractor which also can be replaced by BiGRU because both win comparable performance and better than other sequence models.
\begin{equation}\label{eq2}
\setlength{\abovedisplayskip}{1pt}
\setlength{\belowdisplayskip}{1pt}
\small
  s_1 = BiLSTM(a_1x_1)
\end{equation}
\begin{equation}\label{eq3}
\setlength{\abovedisplayskip}{0pt}
\setlength{\belowdisplayskip}{0pt}
\small
  s_{t-1} = BiLSTM(a_1x_1, ... , a_{t-1}x_{t-1})
\end{equation}
\begin{equation}\label{eq4}
\setlength{\abovedisplayskip}{0pt}
\setlength{\belowdisplayskip}{0pt}
\small
  G_\theta (a_t|s_{t-1};\theta) = sigmoid(Ws_{t-1}+b)
\end{equation}
Following \cite{williams1992simple, sutton2000policy}, temporarily without considering the intermediate reward, the generated policy $G_\theta$ generates a complete sequence to maximize the expected reward $J(\theta)$.
\begin{equation}\label{eq5}
\setlength{\abovedisplayskip}{1pt}
\setlength{\belowdisplayskip}{1pt}
\small
  \mathnormal{J}(\theta) = \mathbb{E}[R_T|s_{t-1}, \theta] = \sum_{x_t}G_\theta(x_t|s_{t-1})Q_{D_\phi}^{G_\theta}(s_{t-1},x_t)
\end{equation}
where $R_T$ is the reward for the whole sequence with $T$ tokens, $Q_{D_\phi}^{G_\theta}(s_{t-1},x_t)$ is the action-value function when the state is $s_{t-1}$, and the action is to choose the input token $x_t$. In this paper, we adopt REINFORCE algorithm \cite{williams1992simple} and consider the discriminator $D_\phi (s_T)$ as the reward to evaluate $Q_{D_\phi}^{G_\theta}(s_{T-1},x_T)$ of the whole sequence, i.e., $Q_{D_\phi}^{G_\theta}(s_{T-1},x_T)=-D_\phi (s_T)$ . This means, the weaker of ability of the discriminator in task 1 to distinguish true and false information is, the higher the $Q_{D_\phi}^{G_\theta}$ value is. When $Q_{D_\phi}^{G_\theta}$ secures the maximum value, the encoding of the state $s_T$ are common features $F_s$.
\begin{equation}\label{eq5.1}
\setlength{\abovedisplayskip}{1pt}
\setlength{\belowdisplayskip}{1pt}
\small
    F_s = s_T \;\; for \; max_{s_T}Q_{D_\phi}^{G_\theta}
\end{equation}
The reward of a finished sequence has been evaluated. Note that the rewards of intermediate states are often of critical importance. Intermediate states determine the quality of the chosen words. Here, owing to the current state t is relevant to $t-1$ states, we employ a roll-out policy $ G_\mu $ to sample the $t$ and $t-1$ tokens $N$ times.
\begin{equation}\label{eq6}
\setlength{\abovedisplayskip}{1pt}
\setlength{\belowdisplayskip}{1pt}
\small
  \left\{ s_{t-1}^1, ... , s_{t-1}^N \right\} = G_\mu(s_{t-1}; N)
\end{equation}
\begin{equation}\label{eq7}
\setlength{\abovedisplayskip}{1pt}
\setlength{\belowdisplayskip}{1pt}
\small
  \left\{ s_t^1, ... , s_t^N \right\} = G_\mu(s_t;N)
\end{equation}
where $s_{t-1}^n=BiLSTM(a_1^n x_1^n, ... ,a_{t-1}^n x_{t-1}^n)$, $G_\mu$ is the copy of the generated policy $G_\theta$. In current state $t$, in order to obtain more precise action value, we rely on above tokens to evaluate $Q_{D_\phi}^{G_\theta} (s_{t-1},x_t)$ by average operation. $Q_{D_\phi}^{G_\theta} (s_{t-1},x_t)$ is formulated as follows:
\begin{equation}\label{eq8}
\setlength{\abovedisplayskip}{1pt}
\setlength{\belowdisplayskip}{1pt}
\scriptsize
\begin{aligned}
  &Q_{D_\phi}^{G_\theta} (s_{t-1},x_t)\! \\
  & =\! \left\{\!\begin{array}{lr}
                                                  \!\frac{1}{N}\sum_{n=1}^{N}\{D_\phi(s_t^n)\!-\!D_\phi(s_{t-1}^n)\} \; s_t^n, s_{t-1}^n \in G_\mu (s_t;N)& \! t < T \\
                                                  \!D_\phi(s_T) & \! t = T
                                                \end{array}
                                                \right.
\end{aligned}
\end{equation}
Additionally, using the discriminator $D_\phi$ as the reward can effectively update and further improve the generated policy $G_\theta$ iteratively. Here, we use the conventional update strategy of discriminator by following the typical rules of GAN \cite{goodfellow2014generative}, $D_\phi$ can be formulated as
\begin{equation}\label{eq9}
\setlength{\abovedisplayskip}{1pt}
\setlength{\belowdisplayskip}{1pt}
\small
L_{adv} = min_\phi -\mathbb{E}_{Y\sim p_{data}}[logD_\phi(s)] - \mathbb{E}_{Y\sim G_\theta}[1-logD_\phi(s)]
\end{equation}
where $s$ denotes the sampling sequence, which is the general representation of $s_{t-1}^n$ without considering time $t$.

Following \cite{sutton2000policy}, when the discriminator is updated, we set out to improve the generator using the policy to maximize the long-term reward. The gradient of the objective function can be derived as
% 公式具体推导：
\begin{equation}\label{eq9.1}
\setlength{\abovedisplayskip}{1pt}
\setlength{\belowdisplayskip}{1pt}
\small
    \begin{aligned}
      \bigtriangledown_\theta \mathnormal{J}(\theta) \\
      & = \bigtriangledown_\theta[\sum_{x_1} G_\theta(x_1|s_0) Q_{D_\phi}^{G_\theta}(s_0,x_1)]\\
      & = \sum_{x_1}[\bigtriangledown_\theta G_\theta(x_1|s_0) Q_{D_\phi}^{G_\theta}(s_0,x_1) + G_\theta(x_1|s_0) \bigtriangledown_\theta Q_{D_\phi}^{G_\theta}(s_0,x_1)] \\
      & = \sum_{x_1} \bigtriangledown_\theta G_\theta(x_1|s_0) Q_{D_\phi}^{G_\theta}(s_0,x_1) \\
      & + \sum_{x_1} G_\theta(x_1|s_0) \bigtriangledown_\theta [\sum_{x_2} G_\theta(x_2|s_1) Q_{D_\phi}^{G_\theta}(s_1,x_2)] \\
      & = \sum_{x_1} \bigtriangledown_\theta G_\theta(x_1|s_0) Q_{D_\phi}^{G_\theta}(s_0,x_1) \\
      & + \sum_{x_1} G_\theta(x_1|s_0) \sum_{x_2}[\bigtriangledown_\theta G_\theta(x_2|s_1) Q_{D_\phi}^{G_\theta}(s_1,x_2) \\
      &     + G_\theta(x_2|s_1) \bigtriangledown_\theta Q_{D_\phi}^{G_\theta}(s_1,x_2)] \\
      & = \sum_{x_1} \bigtriangledown_\theta G_\theta(x_1|s_0) Q_{D_\phi}^{G_\theta}(s_0,x_1) \\
      & + \sum_{s_1} P(s_1|s_0; G_\theta) \sum_{x_2} \bigtriangledown_\theta G_\theta(x_2|s_1) Q_{D_\phi}^{G_\theta}(s_1,x_2) \\
      &     + \sum_{s_2} P(s_2|s_0; G_\theta)\bigtriangledown_\theta \sum_{s_2} G_\theta (x_3|s_2) Q_{D_\phi}^{G_\theta} (s_2, x_3) \\
      & = \sum_{t=1}^{T} \sum_{s_{t-1}} P(s_{t-1}|s_0; G_\theta) \sum_{x_t} \bigtriangledown_\theta G_\theta(x_t|s_{t-1}) Q_{D_\phi}^{G_\theta}(s_{t-1},x_t) \\
      & = \sum_{t=1}^{T} \mathbb{E}_{s_{t-1}\sim G_\theta}[\sum_{x_t}\bigtriangledown_\theta G_\theta(x_t|s_{t-1})Q_{D_\phi}^{G_\theta}(s_{t-1}, x_t)]
    \end{aligned}
\end{equation}
\begin{equation}\label{eq10}
\setlength{\abovedisplayskip}{1pt}
\setlength{\belowdisplayskip}{1pt}
\small
  \bigtriangledown_\theta \mathnormal{J}(\theta) = \mathbb{E}_{s_{t-1}\sim G_\theta}[\sum_{x_t}\bigtriangledown_\theta G_\theta(x_t|s_{t-1})Q_{D_\phi}^{G_\theta}(s_{t-1}, x_t)]
\end{equation}
where the detailed derivation of Eq.(\ref{eq10}) is in Eq.(\ref{eq9.1}). Using likelihood ratios \cite{glynn1990likelihood, sutton2000policy}, we update the policy network with the following gradient:
\begin{equation}\label{eq11}
\setlength{\abovedisplayskip}{1pt}
\setlength{\belowdisplayskip}{1pt}
\small
    \begin{aligned}
      \bigtriangledown_\theta \mathnormal{J}(\theta) \\
      & \simeq \frac{1}{T}\sum_{t=1}^{T}\sum_{x_t}\bigtriangledown_\theta G_\theta(x_t|s_{t-1}) Q_{D_\phi}^{G_\theta}(s_{t-1}, x_t)\\
      & = \frac{1}{T} \sum_{t=1}^{T}\sum_{x_t}G_\theta(x_t|s_{t-1})\bigtriangledown_\theta logG_\theta(x_t|s_{t-1}) Q_{D_\phi}^{G_\theta}(s_{t-1}, x_t) \\
      & = \frac{1}{T} \sum_{t=1}^{T}\mathbb{E}_{x_t \sim G_\theta(x_t|s_{t-1})}\bigtriangledown_\theta logG_\theta(x_t|s_{t-1}) Q_{D_\phi}^{G_\theta}(s_{t-1},x_t)
  \end{aligned}
\end{equation}
\begin{equation}\label{eq12}
\setlength{\abovedisplayskip}{1pt}
\setlength{\belowdisplayskip}{1pt}
\small
  \theta \leftarrow \theta + \varepsilon \bigtriangledown \mathnormal{J}(\theta)
\end{equation}
where $s_{t-1}$ is obtained easily by $G_\theta$. $\varepsilon \in R^+$ denotes learning rate.

% 3.3
\subsection{Task 2 for Multi-classification of Information Credibility}
\subsubsection{Private Feature Extractor}
In Task 2, private features extractor is implemented by BiLSTM. BiLSTM provides complete context information at any positions of inputs for outputs. Specifically, the inputs in this module, the same as Task 1, are the concatenation of word embeddings and the representation of meta-data features. The outputs are called private features.
\begin{equation}\label{eq13}
\setlength{\abovedisplayskip}{1pt}
\setlength{\belowdisplayskip}{1pt}
\small
  P_t = BiLSTM(x_t, P_{t-1}, \phi_p)
\end{equation}
\begin{equation}\label{eq13.1}
\setlength{\abovedisplayskip}{1pt}
\setlength{\belowdisplayskip}{1pt}
\small
  F_p = P_T
\end{equation}
where $x_t$,$P_{t-1}$ represents the inputs in step $t$ and a hidden state in step $t-1$, respectively. $T$ denotes the number of steps in the inputs.

\subsubsection{Multiple Classifier}
Multiple classifier in Task 2 relies on Softmax function to multi-class classification of information credibility. In our model, the inputs are the private features optimized by the following strategies (The details are shown in subsection \ref{strategies}), and the outputs are the results of multi-class classification.
\begin{equation}\label{eq17}
\setlength{\abovedisplayskip}{1pt}
\setlength{\belowdisplayskip}{1pt}
\small
  C(F_p, \phi_c) = softmax(UF_p + b)
\end{equation}
where $U\in R^{d\times d}$ is a learnable parameter and $b\in \mathbb{R}^d$ is a bias. Here, $F_p$ represents the private features.

\textbf{Task 2 Loss.} We adopt cross-entropy algorithm to calculate loss function of Task 2 and train the parameters of the network to minimize it.
\begin{equation}\label{eq18}
\setlength{\abovedisplayskip}{1pt}
\setlength{\belowdisplayskip}{1pt}
\small
  L_{task}(C_i, y_i^2) = -\sum_{i=1}^{N_m}y_i^2log(C_i(F_p,\phi_c))
\end{equation}
where $N_m$ represents the number of samples of the corpus, $y_i$ denotes the label of the sample $i$.

% 3.4
\subsection{Feature Separation Strategies}
\label{strategies}
To obtain better and purer differential features from the private features, we design two strategies in feature separation module: orthogonality constraints are to make common features and the private features independent of each other to reduce the correlations between them; and KL-divergence is to make the common features and the private features more different from each other to boost the diversity of the private features.

\subsubsection{Orthogonality Constraints}
For the first strategy, to retain the independence of common features and private features, we utilize orthogonal constraints to penalize the redundant latent representations and encourage the common and private extractors to encode different aspects of the inputs. Here, we adopt below loss function investigated by Bousmalis et al. \cite{bousmalis2016domain} to obtain better results.
\begin{equation}\label{eq14}
\setlength{\abovedisplayskip}{1pt}
\setlength{\belowdisplayskip}{1pt}
\small
  L_{ind} = ||F_s^TF_p||_F^2
\end{equation}
where $||\;||_F^2$ is the squared Frobenius norm. $F_s^T$ and $F_p$ are two matrices, whose rows are the outputs of common feature extractor and private feature extractor, respectively.

\subsubsection{Negative KL-Divergence}

A foreseeable phenomenon is that orthogonal constraints in the first strategy are difficult to equal zero in practice, which implies there are some correlations between the two types of features. To relieve the correlations and boost the diversity of the private features, we develop the second strategy, namely, negative KL-difference, to measure the differences between the two types of features. $L_{simi}$ denotes the loss produced by the comparison of the common features and the private features via KL-divergence method. $L_{diff}$ as difference measurement has opposite growth trends compared with $L_{simi}$. The calculation needs the private features matrices $F_s$ and the common features matrices $F_p$ to convert into a multi-dimensional vector $S'$ and $P'$, respectively, i.e., $S'=[S'_0,S'_1, ... ,S'_i, ... ,S'_l]$, $P'=[P'_0,P'_1, ... ,P'_i, ... ,P'_l]$, where $l$ represents the size of the multi-dimensional vector $S'$ or $P'$. When $L_{diff}$ is up to maximum finally, the private features optimized contain the maximum number of differential features.
\begin{equation}\label{eq15}
\setlength{\abovedisplayskip}{1pt}
\setlength{\belowdisplayskip}{1pt}
\small
  L_{simi} = \sum_{i}^lS'_ilog(\frac{S'_i}{P'_i})
\end{equation}
\begin{equation}\label{eq16}
\setlength{\abovedisplayskip}{1pt}
\setlength{\belowdisplayskip}{1pt}
\small
  L_{diff} = \frac{1}{L{simi}}
\end{equation}
% 3.6
\subsection{Total Loss}
According to above all steps, to ensure the effective integration and synergy of the two tasks, we put together all loss mentioned above to get the total loss $L$.
\begin{equation}\label{eq19}
\setlength{\abovedisplayskip}{1pt}
\setlength{\belowdisplayskip}{1pt}
\small
  L = L_{task} + \alpha L_{adv} + \beta L_{ind} + \gamma L_{diff}
\end{equation}
Algorithm 1 formally provides specific steps of an epoch, where specific formulas are described in detail in section \ref{sec:3}.

% 添加算法流程图
% 表 算法流程图
% 控制行高度
\linespread{1}
\begin{table}
\small
\center
% \caption{}
  \label{algorithm1}
  \setlength{\tabcolsep}{0mm}{
  \begin{tabular}{rl}
    \toprule
    \multicolumn{2}{l}{\textbf{Algorithm 1: The ANSP model}}\\
    \midrule
    \multicolumn{2}{l}{\textbf{Require:} Corpus $\mathbb{X}=\{(x_i, y_i^1, y_i^2)\}_{i=1}^N$; generator policy $G_\theta$;} \\
    \multicolumn{2}{l}{roll-out policy $G_\mu$; discriminator $D_\phi$; hyperparameter $\alpha$, $\beta$, $\gamma$} \\
    1: & Initialize $G_\theta$, $D_\phi$ with random weights $\theta$, $\phi$\\
    2: & Pre-train $G_\theta$ using the maximum likelihood estimation on $\mathbb{X}$\\
    3: & Pre-train $D_\phi$ via minimizing the cross entropy\\
    4: & $\mu \leftarrow \theta$\\
    5: & $Loss = 0$ \\
    \textbf{6}: & \textbf{Repeat}\\
    \textbf{7}: &$\big|$ \quad \textbf{for}  $i=1$ to $N$ \textbf{do}\\
    8: &$\big|$ \quad$\big|$\quad //Adversarial networks guided by RL\\
    \textbf{9}: &$\big|$ \quad$\big|$\quad \textbf{for} g-steps \textbf{do}\\
    10:&$\big|$ \quad$\big|$\quad$\big|$\quad using $G_\theta$ to generate a sequence and the length is $T$\\
    \textbf{11}:&$\big|$ \quad$\big|$\quad$\big|$\quad \textbf{for} $t$  in  $1:T$ \textbf{do}\\
    12:&$\big|$ \quad$\big|$\quad$\big|$\quad$\big|$\quad compute $Q(s_{t-1}, x_t)$ by eq.(\ref{eq8})\\
    \textbf{13}:&$\big|$ \quad$\big|$\quad$\big|$\quad \textbf{end for}\\
    14:&$\big|$ \quad$\big|$\quad$\big|$\quad update generator via policy gradient eq.(\ref{eq12})\\
    \textbf{15}:&$\big|$ \quad$\big|$\quad \textbf{end for}\\
    \textbf{16}:&$\big|$ \quad$\big|$\quad\textbf{for} d-steps \textbf{do}\\
    17:&$\big|$ \quad$\big|$\quad$\big|$\quad use given two kinds of labels examples\\
    18:&$\big|$ \quad$\big|$\quad$\big|$\quad train discriminator $D_\phi$ by eq.(\ref{eq9})\\
    19:&$\big|$ \quad$\big|$\quad$\big|$\quad Get adversarial loss $L_{adv}$\\
    20:&$\big|$ \quad$\big|$\quad$\big|$\quad $Loss\;+=\;\alpha L_{adv}$\\
    \textbf{21}:&$\big|$ \quad$\big|$\quad \textbf{end for}\\
    \textbf{22}:&$\big|$ \quad$\big|$\quad \textbf{Extract common features} $F_s$ by eq.(\ref{eq5.1}) \\
    \textbf{23}:&$\big|$ \quad$\big|$\quad \textbf{Extract private features} $P(x_i, \phi_p)$ by eq.(\ref{eq13}) \\
    24:&$\big|$ \quad$\big|$\quad\quad $F_p = P(x_i,\phi_p)$\\
    \textbf{25}:&$\big|$ \quad$\big|$\quad \textbf{Capture independence features} to get\\
    26:&$\big|$ \quad$\big|$\quad\quad independence loss by orthogonality constraints\\
    27:&$\big|$ \quad$\big|$\quad\quad $L_{ind}(F_s, F_p)$ by eq.(\ref{eq14})\\
    28:&$\big|$ \quad$\big|$\quad\quad $Loss\; +=\; \beta L_{ind}(F_s, F_p)$\\
    \textbf{29}:&$\big|$ \quad$\big|$\quad \textbf{Capture differential features} to get differential loss \\
    30:&$\big|$ \quad$\big|$\quad\quad $L_{diff}(F_s, F_p)$ by eq.(\ref{eq15}), eq.(\ref{eq16})\\
    31:&$\big|$ \quad$\big|$\quad\quad $Loss\; +=\; \gamma L_{diff}(F_s, F_p)$\\
    \textbf{32}:&$\big|$ \quad$\big|$\quad \textbf{ Train multiple classifier} to obtain task loss\\
    33:&$\big|$ \quad$\big|$\quad\quad $L_{task}(C_i, y_i^2)$ by eq.(\ref{eq18})\\
    34:&$\big|$ \quad$\big|$\quad\quad $Loss\; +=\; L_{task}(C_i, y_i^2)$\\
    35:&$\big|$ \quad$\big|$\quad Update the parameters of $F_p$, $F_s$, $C_i$ using $\nabla$Loss\\
    \textbf{36}:&$\big|$ \quad \textbf{end for}\\
    37:&$\big|$ $\mu \leftarrow \theta$\\
    \textbf{38}:&\textbf{until} convergence \\
    \bottomrule
  \end{tabular}
}
\end{table}

% 4. 实验
\section{Experiments}
\label{sec:4}
In this work, we first systematically evaluate the performance of ANSP on six-label LIAR and Weibo datasets around the following perspectives: the effectiveness of the model, the performance of ANSP based on different types of information and different inputs, model ablation, and the scalability of ANSP. Then, we validate the robustness of the model on four-label Twitter16. Finally, the limitations to the model are analyzed according to the above experimental results.

\subsection{Datasets}
\label{sec:4.1}
We use six-label LIAR \cite{wang2017liar} and Weibo collected by us to evaluate the effectiveness of ANSP and utilize four-label Twitter16 \cite{ma2017detect} to testify the robustness of the model. The details of the three datasets are described as follows:

LIAR includes 12.8K labeled fake news from politifact.com\footnote{https://www.politifact.com} and every piece of fake news has seven types of meta-data: subject, context/venue, speaker, speaker's job title, state, party affiliation, and prior history. The dataset is classified based on six types, i.e., pants on fire, false, barely-true, half-true, mostly-true, and true. The distribution of labels in LIAR is respectively well-balanced and the number of entries in every type of labels is about 2000.

Weibo is a Chinese false information dataset collected from Sina Weibo Community Management Center\footnote{https://service.account.weibo.com}. It includes 18K labeled microblogs, where a microblog consists of text content and eight meta-data elements: the number of reposts (RP), the quantity of thumbs (TU), the number of comments (CM), the location (LOC) of the microblog, the level (LV) of the user, following (FI) of the user, followers (FL) of the user, and number of times the user releases messages (MT) in a day. The classification of Weibo, the same as LIAR, is also based on six types. Every microblog is labeled by 20 annotators invited and we choose the highest votes as the final annotation results in terms of credibility annotation criteria. Table \ref{tab:1} gives some snapshot information of two datasets.

Twitter16 contains conversation threads associated with different newsworthy events. The conservation consists of a group of widespread source tweets, i.e., retweet and responses, expressing their opinions towards the claim contained in the source tweet. one tweet contains more than 20 types of meta-data: username, screen name, gender, reposts count, bi-followers count, followers count, friends count, attitudes count, statuses count, favourites count, comments count, user avatar, city, province, user location, picture, parent, is verified, geo, user created at, etc. The dataset is divided into four categories: non-rumor (NR), false rumor (FR), true rumor (TR), and unverified rumor (UR). Table 2 shows the statistics of twitter16.

% 表1
\begin{table}
\center
\small
  \caption{The statistics of LIAR and Weibo datasets}
  \label{tab:1}
  \setlength{\tabcolsep}{5.5mm}{
  \begin{tabular}{lcc}
    \toprule
    &LIAR&Weibo\\
    \midrule
    Pants on fire & 1050 & 3041 \\
    False & 2511 & 3035 \\
    Barely-true & 2108 & 3056 \\
    Half-true & 2638 & 3048 \\
    Mostly-true & 2466 & 3078 \\
    True & 2063 & 3020 \\
  \bottomrule
\end{tabular}}
\end{table}

\iffalse
% 表 Twitter16统计 - 横着的统计
\begin{table}
\center
\small
  \caption{The statistics of Twitter16}
  \label{tab:横着的表格}
  \setlength{\tabcolsep}{5mm}{
  \begin{tabular}{lcccccc}
    \toprule
    \# of users & \# of source tweets & \# of threads & \# of non-rumors (NR) & \# of false rumors (FR) & \# of true rumors (TR) & \# of unverified rumors (UR) \\
    \midrule
    173,487 & 818 & 204,820 & 205 & 205 & 205 & 203 \\
  \bottomrule
\end{tabular}
}
\end{table}
\fi

% 表 Twitter16统计 - 竖着的统计
\begin{table}
\center
\small
  \caption{The statistics of Twitter16}
  \label{tab:twitter16-statistics}
  \setlength{\tabcolsep}{5mm}{
  \begin{tabular}{lc}
    \toprule
    \ Parameters & Number \\
    \hline
    \# of users & 173,487 \\
    \# of source tweets & 818 \\
    \# of threads & 204,820 \\
    \# of non-rumors (NR) & 205 \\
    \# of false rumors (FR) & 205 \\
    \# of true rumors (TR) & 205 \\
    \# of unverified rumors (UR) & 203 \\
  \bottomrule
\end{tabular}}
\end{table}

% 表 parameters
\begin{table}
\center
\small
  \caption{The parameters of ANSP on different datasets}
  \label{tab:parameters}
  \setlength{\tabcolsep}{2.5mm}{
  \begin{tabular}{rccccc}
    \toprule
     & \textbf{Dropout} & \textbf{LRate} & \textbf{$\alpha$} & \textbf{$\beta$} & \textbf{$\gamma$} \\
    \midrule
    LIAR & 0.6 & 0.01 & 0.5 & 0.3 & 0.2 \\
    Weibo & 0.5 & 0.01 & 0.6 & 0.2 & 0.2\\
    Twitter16 & 0.5 & 0.01 & 0.6 & 0.15 & 0.25\\
  \bottomrule
\end{tabular}
}
\end{table}

In our model, Task 1 requires two types of labels and Task 2 needs multi-class labels. LIAR and Weibo can provide directly six kinds of labels for Task 2, while they are unable to supply two kinds of labels for Task 1. Remarkably, since the input features in Task 2 desire to reduce the common features of true and false information (a.k.a. the goal of Task 1), the labels of Task 1 must cover the labels of Task 2. Therefore, the six kinds of labels are merged into two kinds of labels for Task 1 in terms of the degree of credibility. We integrate half-true, mostly-true, and true into the true label. The remaining three types are integrated into the false label. Similarly, on Twitter16, Task 1 is used to classify true and false rumors and task 2 is used to evaluate four types of information.

\textbf{Datasets Partitioned:} The three datasets are partitioned randomly into training set, development set and testing set with the proportion of 80\%, 10\%, and 10\%.

\textbf{Evaluation Metrics:} To make fair comparisons for our model and baseline models, we select accuracy (\textbf{A}), precision (\textbf{P}), recall (\textbf{R}), F1-score (\textbf{F1}) as metrics on LIAR and Weibo. On Twitter16, we use \textbf{A} to evaluate overall performance and use \textbf{F1} to evaluate the performance of models on each type of information.

% 表2
\linespread{1.5}
%\begin{sidewaystable}
%\begin{adjustbox}{angle=90}[h]
 \begin{table*}
 \linespread{1.2}
\center
\small
  \caption{The performance evaluation of ANSP on LIAR and Weibo}
  \label{tab:2}
  \setlength{\tabcolsep}{4mm}{
  \begin{tabular}{r|cccc|ccccc}
    \toprule
    &\multicolumn{4}{c|}{LIAR} & \multicolumn{4}{c} {Weibo} \\
    \midrule
    & \textbf{A} & \textbf{P} & \textbf{R} & \textbf{F1} & \textbf{A} & \textbf{P} & \textbf{R} & \textbf{F1}\\ 
    LR & 0.257 & 0.252 & 0.236 & 0.244 &  0.281 & 0.296 & 0.270 & 0.282\\
    SVMs & 0.258 & 0.258 & 0.233 & 0.245 & 0.305 & 0.315 & 0.291 & 0.303\\
    CNN & 0.260 & 0.278 & 0.249 & 0.263 & 0.321 & 0.332 & 0.313 & 0.322\\
    LSTM & 0.241 & 0.254 & 0.227 & 0.240 & 0.306 & 0.324 & 0.288 & 0.305\\
    BiLSTM & 0.223 & 0.243 & 0.221 & 0.231 & 0.297 & 0.316 & 0.281 & 0.297\\
    Hybrid-CNN & 0.277 & 0.286 & 0.258 & 0.271 & 0.373 & 0.392 & 0.357 & 0.374\\
    Multi-Attention & 0.407 & 0.415 & 0.379 & 0.396 & 0.432 & 0.452 & 0.401 & 0.425\\
    %\specialrule{0pt}{1pt}{1pt}
    %\rowcolor{mygray} % 设置下一行的背景颜色为灰色
    \textbf{ANSP} & \textbf{0.428} & \textbf{0.423} & \textbf{0.398} & \textbf{0.410} & \textbf{0.463} & \textbf{0.490} & \textbf{0.437} & \textbf{0.462}\\

  \bottomrule
\end{tabular}}
\end{table*}
%\end{adjustbox}
%\end{sidewaystable}

% 4.3
\subsection{Model Setup}

\textbf{Hyperparameters:} In our experiments, we use pre-trained 200-dimensional word2vec embeddings from English and Chinese Wikipedia \cite{Mikolov2013Efficient} to represent text content of the datasets, respectively. For each task, we strictly turn all the hyperparameters on the validation dataset, and we achieve the best performance via a small grid search. Finally, the dropout keep probabilities (Dropout), the learning rate (LRate) and hyperparameters are learned on different datasets and shown in Table \ref{tab:parameters}.

\textbf{Features:} Besides text features, several non-sparse features of two datasets, meta-data features, like subject, speaker, job, state, party, etc. are participated in our model as important credibility features. We employ 20-dimensional vector to represent single meta-data feature, the parts that vectors of one feature are less than 20-dimensional will be filled zero.

Additionally, our all experiments use open\-source framework Tensorflow \footnote{https://www.tensorflow.org/} on Linux CentOS 7.6.1810 system with Intel Xeon E5-2620 CPU (64 G memory) and NVIDIA TITAN Xp GPU (12G graphics memory). The time cost of ANSP on three datasets, i.e., LIAR, Weibo, and Twitter16 are 17750, 24540, 32380 seconds for 30 epochs, respectively. The core code is published in github \footnote{https://github.com/wuxiaoxiaoer}.

\subsection{Performance Evaluation}
\label{sub4performanceEva}
Table \ref{tab:2} describes the evaluation results on LIAR and Weibo datasets compared with the following baseline methods:

\textbf{LR:} Although Logistic Regression (LR) is a regression model, it is a classical and effective classification model.

\textbf{SVM:} Support Vector Machines (SVM) \cite{Crammer2001On} model is originally designed for binary classification. Certainly, in this work, we use multiple SVM to achieve multiple-class classification.

\textbf{CNN:} Convolutional Neural Network model (CNN) \cite{Kim2014Convolutional} uses shared convolutional kernel to easily handle high-dimensional data and it is an excellent classification model, especially in image processing.

\textbf{LSTM:} Long Short-Term Memory (LSTM) \cite{hochreiter1997long} is a sequence model that is elaborated to prevent the gradient explosion.

\textbf{BiLSTM:} Bi-directional Long Short-Term Memory networks model (Bi-LSTM) \cite{Zhang2014Sequential} not only prevents the gradient explosion but also captures context information than LSTM model.

\textbf{Hybrid-CNN:} Hybrid-CNN model proposed by Wang \cite{wang2017liar} consists of a convolutional layer and a bi-directional LSTM layer, which integrates meta-data features and text features to detect fake news.

\textbf{Multi-Attention:} Multi-Attention hybrid model explored by Long et al. \cite{long2017fake} focuses on multiple perspectives of speaker profiles to capture valuable features for validating the credibility of news articles.

From Table \ref{tab:2}, we observe that ANSP achieves the superior performance compared with the baseline methods. In details, the model has a 17\% average improvement compared with LR and SVMs because our model with neural networks can fully exploit deep semantic representations of text. The model achieves the state-of-the-art performance on two datasets with 42.8\%, and 46.3\% overall accuracy respectively, surpassing CNN, LSTM, and BiLSTM by a significant margin of 16.8\%, and 14.2\%. This suggests that private features and common features captured by our model based on multi-task learning are effective for credibility evaluation. The model outperforms attention-based method with 2.1\%, 3.1\% improvements. This is because our model reduces common, irrelevant-type features from private features and focuses on the differential credibility features.

% 图5 (c) - (d)
\begin{figure*}
\centering
% 图5 (a) - (b)
\subfigure[Hybrid-CNN on LIAR (A=27.7\%)]{
    \begin{minipage}[b]{0.3\textwidth}
    \includegraphics[width=1\textwidth]{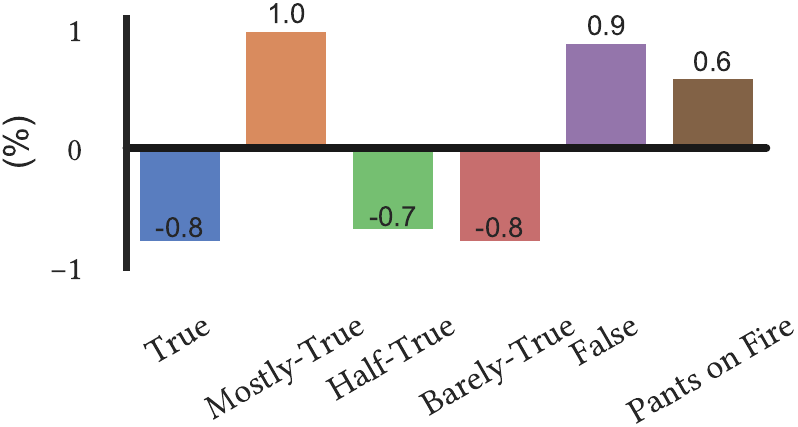}
    \end{minipage}
}
% 子图2： 图5(b)
\subfigure[Hybrid-CNN on Weibo (A=37.3\%)]{
    \begin{minipage}[b]{0.3\textwidth}
    \includegraphics[width=1\textwidth]{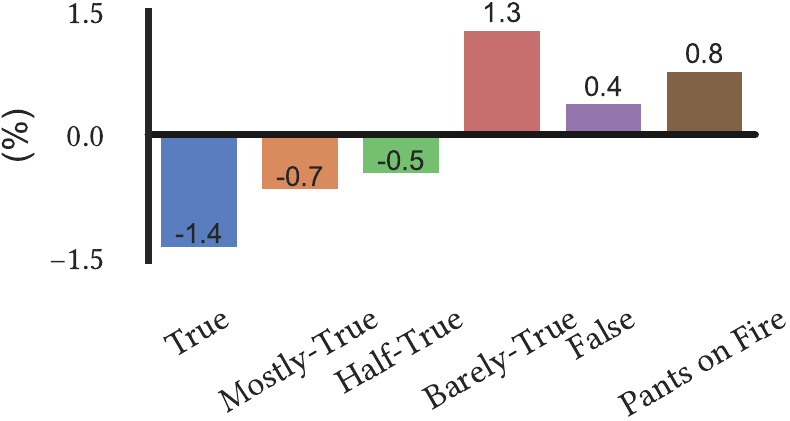}
    \end{minipage}
}
% 子图3： 图5(c)
\subfigure[Multi-Attention on LIAR (A=40.7\%)]{
    \begin{minipage}[b]{0.3\textwidth}
    \includegraphics[width=1\textwidth]{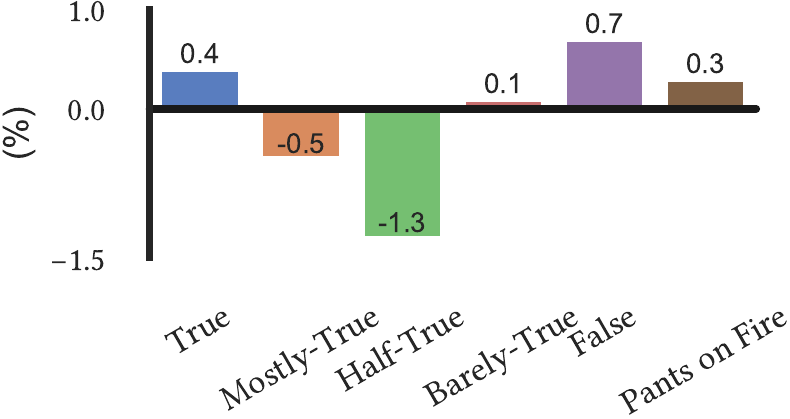}
    \end{minipage}
}

% 子图4： 图5(d)
\subfigure[Multi-Attention on Weibo (A=43.2\%)]{
    \begin{minipage}[b]{0.3\textwidth}
    \includegraphics[width=1\textwidth]{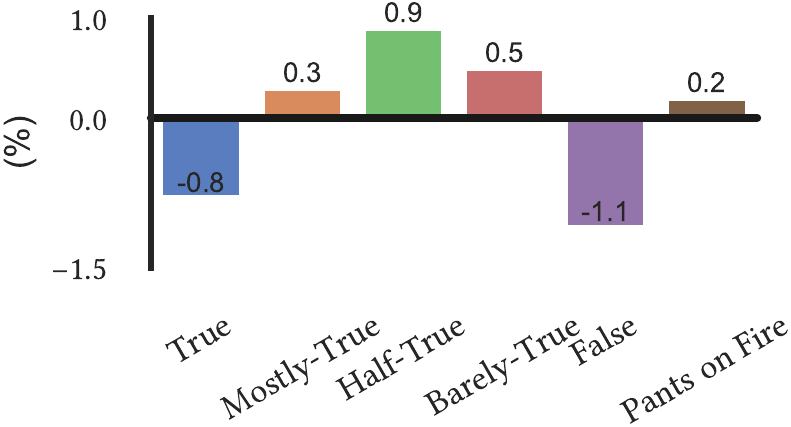}
    \end{minipage}
}
% 子图3： 图5(e)
\subfigure[ANSP on LIAR (A=42.8\%)]{
    \begin{minipage}[b]{0.3\textwidth}
    \includegraphics[width=1\textwidth]{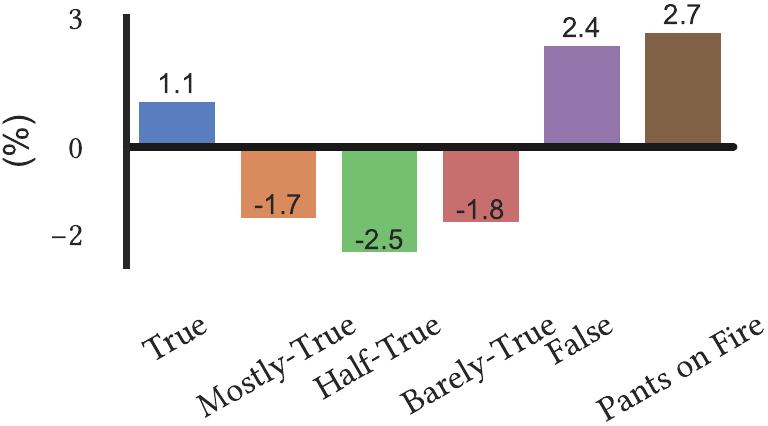}
    \end{minipage}
}
% 子图4： 图5(f)
\subfigure[ANSP on Weibo (A=46.3\%)]{
    \begin{minipage}[b]{0.3\textwidth}
    \includegraphics[width=1\textwidth]{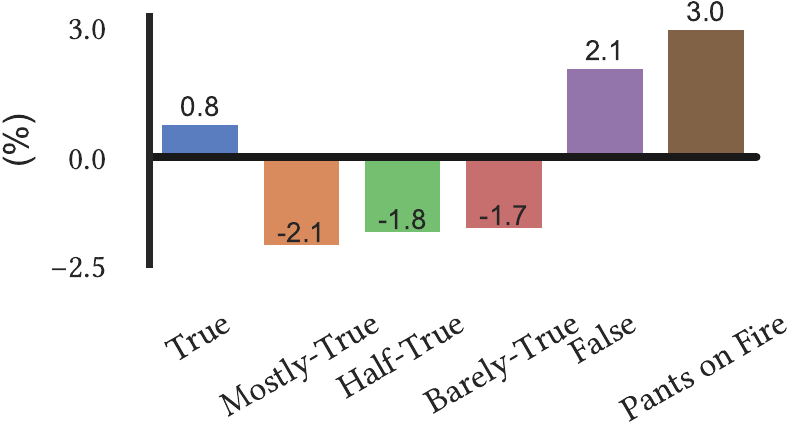}
    \end{minipage}
}
% 总标题
\caption{Accuracy of models based on different types of information. Every bar in the subfigures denotes the difference between the performance of a method in each category of information on a dataset and the overall performance of the method. As a concrete example, In Figure (a), the bar with category `True' is -0.8\%, which represents that the performance of Hybrid-CNN in category `True' on LIAR is 26.9\% (i.e., 27.7\%-0.8\%=26.9\%).} \label{fig:diffr-types}
\end{figure*}

\subsection{Performance of ANSP on Different Types of Information}
\label{sub4detectionAccuracy}
Because we reduced common features of true and false information from private features, this is to say that captured differential private features include more such features that are beneficial to classify these two pure types of information: ``true'', ``false'', instead of ``half true'', ``barely true'', etc., in LIAR and Weibo datasets. Therefore, if captured differential features exist, ANSP will achieve better performance on the types ``true'' and ``false'' than the other types. We analyze the accuracy of models based on different types of information on LIAR and Weibo, as shown in Figure \ref{fig:diffr-types}. We can get the following observations:

\begin{itemize}
    \item The accuracy of ANSP on extreme types ``false" and ``pant on fire" is obviously higher than the other types, while no similar results have been found in other models. And ANSP achieves certain improvements than other models on the type `true'. These two observations are enough to demonstrate that differential features captured by ANSP exist and are effective to classify the two extreme types of information, which might offer a feasible solution for multi-threshold and hierarchical detection and monitoring of fake news.
    \item The results of Hybrid-CNN and Multi-Attention models on six types of information are no obvious differences, i.e., fluctuating around 1\%, which indicates these models do not capture particularly prominent features for specific types of information. It also shows that no type of information be easier recognized if it is not specially processed.
    \item ANSP achieves lower performance on types ``mostly true'', ``half true'', and ``barely true'' than other types of information, i.e., below overall performance about 2\%, which illustrates that one of the limitations for our model is that our model is difficult to identify half-true types of information effectively.
\end{itemize}

% 注释 -开始
\iffalse
% 表5 不同类型下的性能
\begin{table*}

\center
\scriptsize
  \caption{Accuracy of models based on different types of information on Weibo}
  \label{tab:5}
  \setlength{\tabcolsep}{1mm}{
      \begin{tabular}{l|ccc|ccc}
        \toprule
         & \multicolumn{3}{c|}{LIAR} &  \multicolumn{3}{c}{Weibo} \\
         \cline{2-4} \cline{5-7}
         & \textbf{Hybrid-CNN} & \textbf{Multi-Attention} & \textbf{ANSP} & \textbf{Hybrid-CNN} & \textbf{Multi-Attention} & \textbf{ANSP} \\
        \midrule
        True &$0.269_{(-0.8\%)}$ & $0.411_{(+0.4\%)}$ & $0.449_{(+1.1\%)}$  &$0.359_{(-1.4\%)}$ & $0.424_{(-0.8\%)}$ & $0.471_{(+0.8\%)}$\\
        Mostly-True &$0.287_{(+1.0\%)}$ & $0.402_{(-0.5\%)}$ & $0.411_{(-1.7\%)}$  &$0.366_{(-0.7\%)}$ & $0.435_{(+0.3\%)}$ & $0.442_{(-2.1\%)}$\\
        Half-True &$0.270_{(-0.7\%)}$ & $0.394_{(-1.3\%)}$ & $0.403_{(-2.5\%)}$  &$0.368_{(-0.5\%)}$ & $0.441_{(+0.9\%)}$ & $0.445_{(-1.8\%)}$\\
        Barely-True &$0.269_{(-0.8\%)}$ & $0.408_{(+0.1\%)}$ & $0.410_{(-1.8\%)}$  &$0.386_{(+1.3\%)}$ & $0.437_{(+0.5\%)}$ & $0.446_{(-1.7\%)}$\\
        False &$0.286_{(+0.9\%)}$ & $0.414_{(+0.7\%)}$ & $0.452_{(\textbf{+2.4\%})}$  &$0.377_{(+0.4\%)}$ & $0.421_{(-1.1\%)}$ & $0.484_{(\textbf{+2.1\%})}$\\
        Pants on Fire &$0.283_{(+0.6\%)}$ & $0.410_{(+0.3\%)}$ & $0.455_{(\textbf{+2.7\%})}$  &$0.383_{(+0.8\%)}$ & $0.434_{(+0.2\%)}$ & $0.493_{(\textbf{+3.0\%})}$\\
      \bottomrule
      \end{tabular}
  }
\end{table*}
\fi
% 注释 -结束

% 图7
\begin{figure*}
  \centering
  \subfigure[The performance of the model on LIAR]{
    \begin{minipage}[b]{0.4\textwidth}
      \includegraphics[width=1\textwidth]{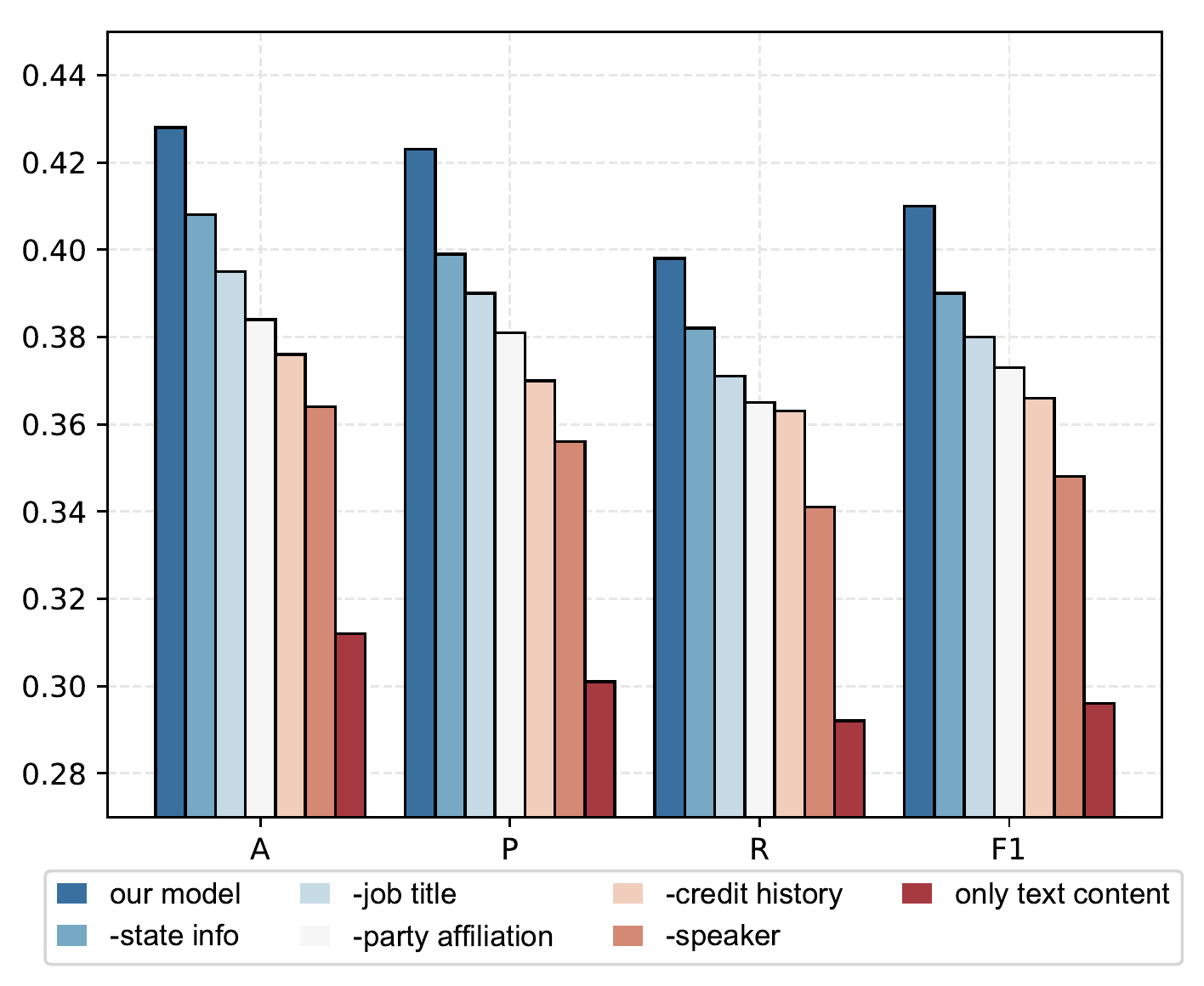}
    \end{minipage}
  }
  \subfigure[The performance of the model on Weibo]{
    \begin{minipage}[b]{0.4\textwidth}
      \includegraphics[width=1\textwidth]{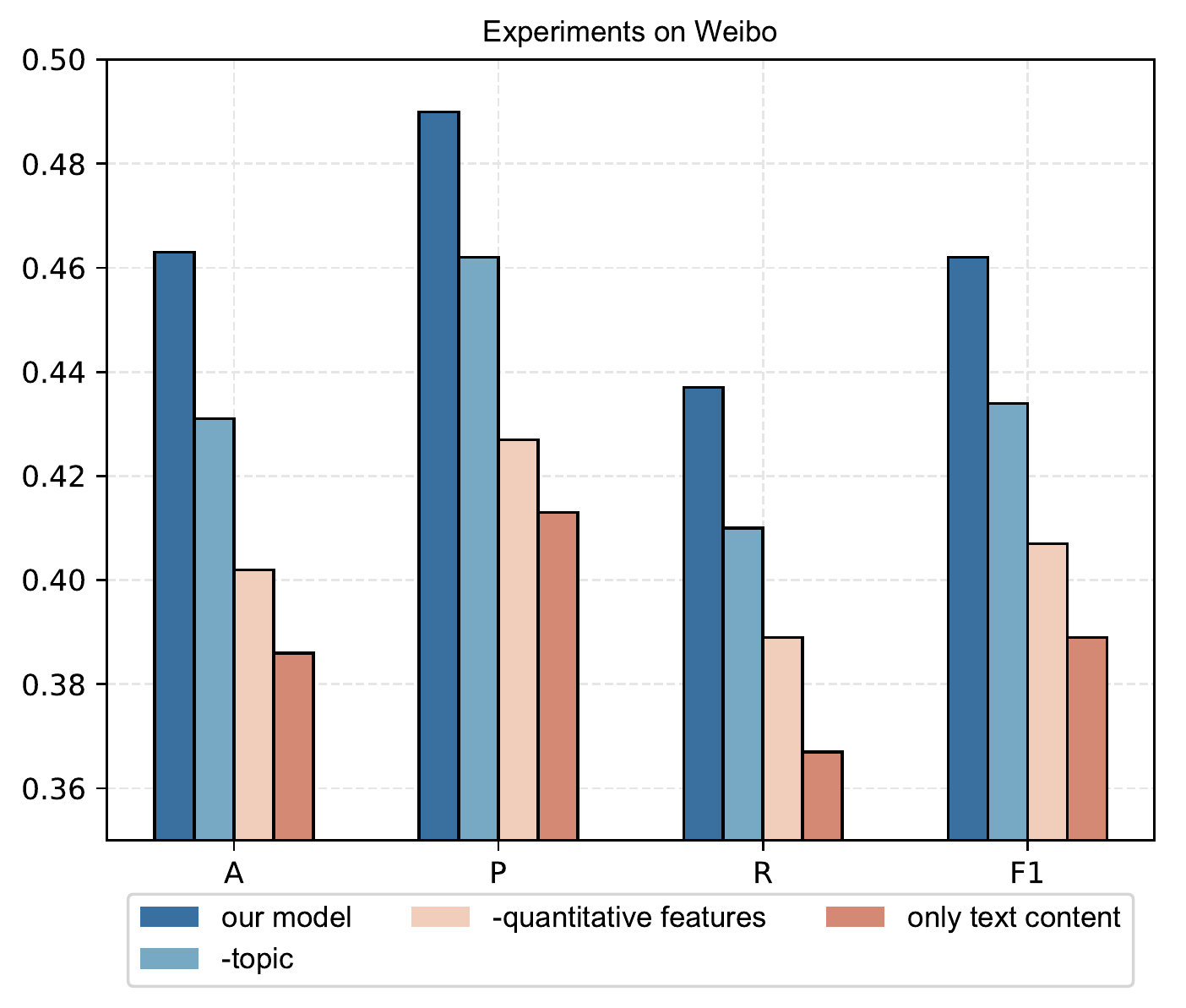}
    \end{minipage}
  }
  \caption{Performance vs. Different features as inputs of the model on different dataset}\label{fig:7}
\end{figure*}

\subsection{Performance of ANSP on Different Inputs}
We further investigate which input features contain much more differential features, we build different input features to exam the performance of ANSP on LIAR and Weibo.

\textbf{all features} refer to the concatenation of text content and all types of meta-data features. \textbf{only text content} means that the model only uses text content as input features. On LIAR, \textbf{-speaker} represents that the model only removes speaker features from \textbf{all features} as input features. Accordingly, \textbf{-job title}, \textbf{-credit history}, \textbf{-state info}, and \textbf{-party affiliation} are removed from \textbf{all features} as inputs features of the model, respectively. On Weibo, \textbf{-topics} and \textbf{-quantitative features} denote that the model only removes topics and quantitative features as input features, respectively. Additionally, \textbf{our model} uses \textbf{all features} to evaluate credibility.

From Figure \ref{fig:7}(a), on LIAR, we observe that methods of removing different features separately achieve different performances. Especially, no considering temporarily \textbf{only text content} method, when only one type of feature is deleted, \textbf{-speaker} method obtains a lower accuracy. This illustrates that speaker features cover more differential credibility features than other features and to a certain extent we can recognize the credibility of what he/she said according to the credibility of his/her profile. Additionally, \textbf{only text content} method achieves the lowest performance, which indicates that it is unwise to only consider text features for credibility evaluation while ignoring meta-data features.

From Figure \ref{fig:7}(b), experimental results on Weibo are similar to the results on LIAR. The methods achieve similar performance differences by removing different features as the inputs of the model. Specifically, \textbf{-quantitative features} method achieves relatively weaker performance, which shows that user interbehavior on Weibo are perfectly valid credibility features in contrast with other features, while also including more differential features.

By comparing Figure \ref{fig:7}(a) and (b), experimental results demonstrate that feature selection has great influences on information credibility evaluation. Meanwhile, we find that the concatenation of text features and meta-data features organically is more beneficial to evaluate credibility than single type of features. Additionally, the results of Figure \ref{fig:7}(b) are obviously higher than Figure \ref{fig:7}(a)'s, which may reveal that the task of credibility evaluation is not only related to different input features, but also to linguistic characteristics between English and Chinese.

\iffalse
% 表  model ablation:
\begin{table}
\linespread{1.2}
\center
\small
  \caption{The performance evaluation of different parts of the model}
  \label{tab:3}
  \setlength{\tabcolsep}{0.2mm}{
  \begin{tabular}{rcccccccc}
    \toprule
    & \multicolumn{4}{|c|}{LIAR} & \multicolumn{4}{c} {Weibo} \\
    \midrule
    & \textbf{A} & \textbf{P} & \textbf{R} & \textbf{F1} & \textbf{A} & \textbf{P} & \textbf{R} & \textbf{F1}\\ \hline
    basic & 0.241 & 0.254 & 0.227 & 0.240 & 0.306 & 0.324 & 0.288 & 0.305\\
    basic+Adv & 0.358 & 0.366 & 0.341 & 0.353 & 0.398 & 0.413 & 0.386 & 0.399\\
    basic+Adv+Ind & 0.392 & 0.403 & 0.387 & 0.395 & 0.437 & 0.449 & 0.424 & 0.436\\
    basic+Adv+Ind+Diff & 0.428 & 0.423 & 0.398 & 0.410 & 0.463 & 0.490 & 0.437 & 0.462\\
  \bottomrule
\end{tabular}
}
\end{table}
\fi

% 表  model ablation:
\begin{table}
\linespread{1.5}
\center
\small
  \caption{The performance evaluation of different parts of the model}
  \label{tab:3}
  \setlength{\tabcolsep}{1.5mm}{
  \begin{tabular}{llcccc}
    \midrule
    & & \textbf{A} & \textbf{P} & \textbf{R} & \textbf{F1} \\ \hline
    \multirow{4}{*}{LIAR} & basic & 0.241 & 0.254 & 0.227 & 0.240 \\
    & basic+Adv & 0.358 & 0.366 & 0.341 & 0.353 \\
    & basic+Adv+Ind & 0.392 & 0.403 & 0.387 & 0.395 \\
    & basic+Adv+Ind+Diff & 0.428 & 0.423 & 0.398 & 0.410 \\ \hline
    
    \multirow{4}{*}{Weibo} & basic & 0.306 & 0.324 & 0.288 & 0.305 \\
    & basic+Adv & 0.398 & 0.413 & 0.386 & 0.399 \\
    & basic+Adv+Ind & 0.437 & 0.449 & 0.424 & 0.436 \\
    & basic+Adv+Ind+Diff & 0.463 & 0.490 & 0.437 & 0.462 \\
  \bottomrule
\end{tabular}
}
\end{table}

% 表4
\begin{table}
\center
\scriptsize
  \caption{The performance of different methods for common features extraction}
  \label{tab:4}
  \setlength{\tabcolsep}{3mm}{
  \begin{tabular}{rcccc}
    \toprule
    Adversarial Networks & \textbf{A} & \textbf{P} & \textbf{R} & \textbf{F1} \\
    \midrule
    with reinforcement learning &0.463 & 0.490 & 0.437 & 0.462\\
    without reinforcement learning&0.426 & 0.447 & 0.384 & 0.413\\
  \bottomrule
\end{tabular}
}
\end{table}

% 图5 (c) - (d)
\begin{figure*}
\centering
% 图5 (a) - (b)
\subfigure[Accuracy]{
    \begin{minipage}[b]{0.23\textwidth}
    \includegraphics[width=1\textwidth]{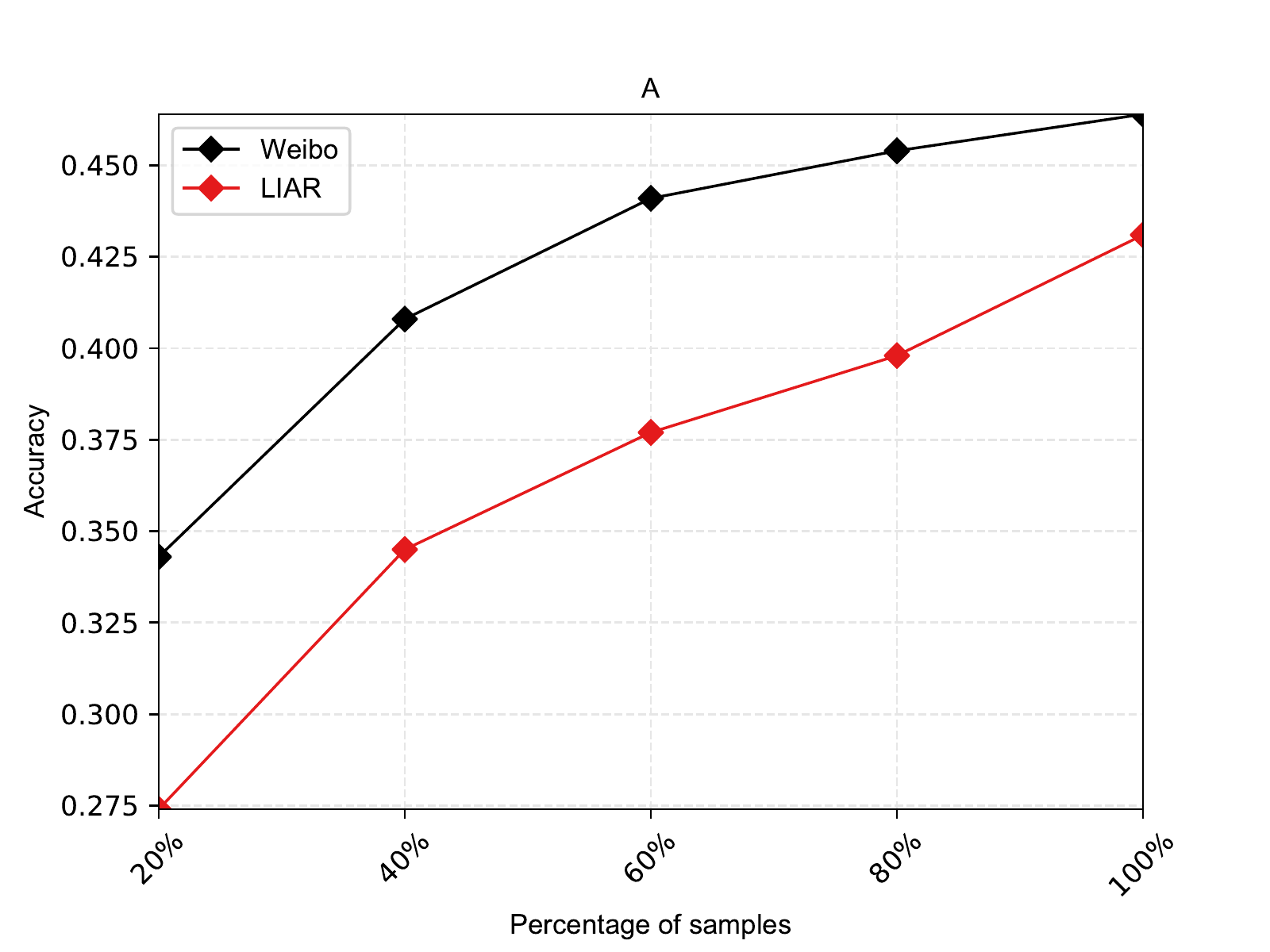}
    \end{minipage}
}
% 子图2： 图5(b)
\subfigure[Precision]{
    \begin{minipage}[b]{0.23\textwidth}
    \includegraphics[width=1\textwidth]{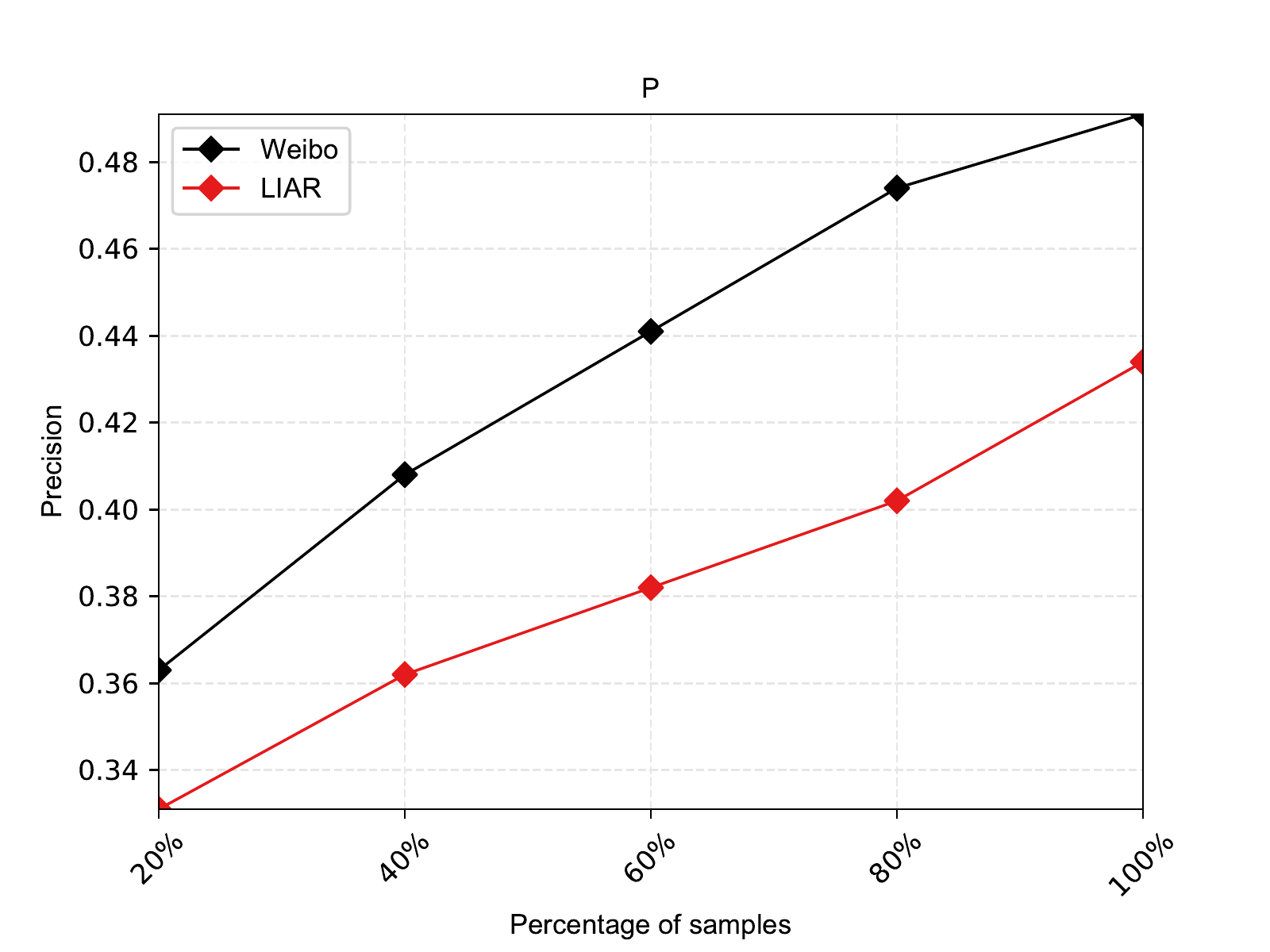}
    \end{minipage}
}
% 子图3： 图5(c)
\subfigure[Recall]{
    \begin{minipage}[b]{0.23\textwidth}
    \includegraphics[width=1\textwidth]{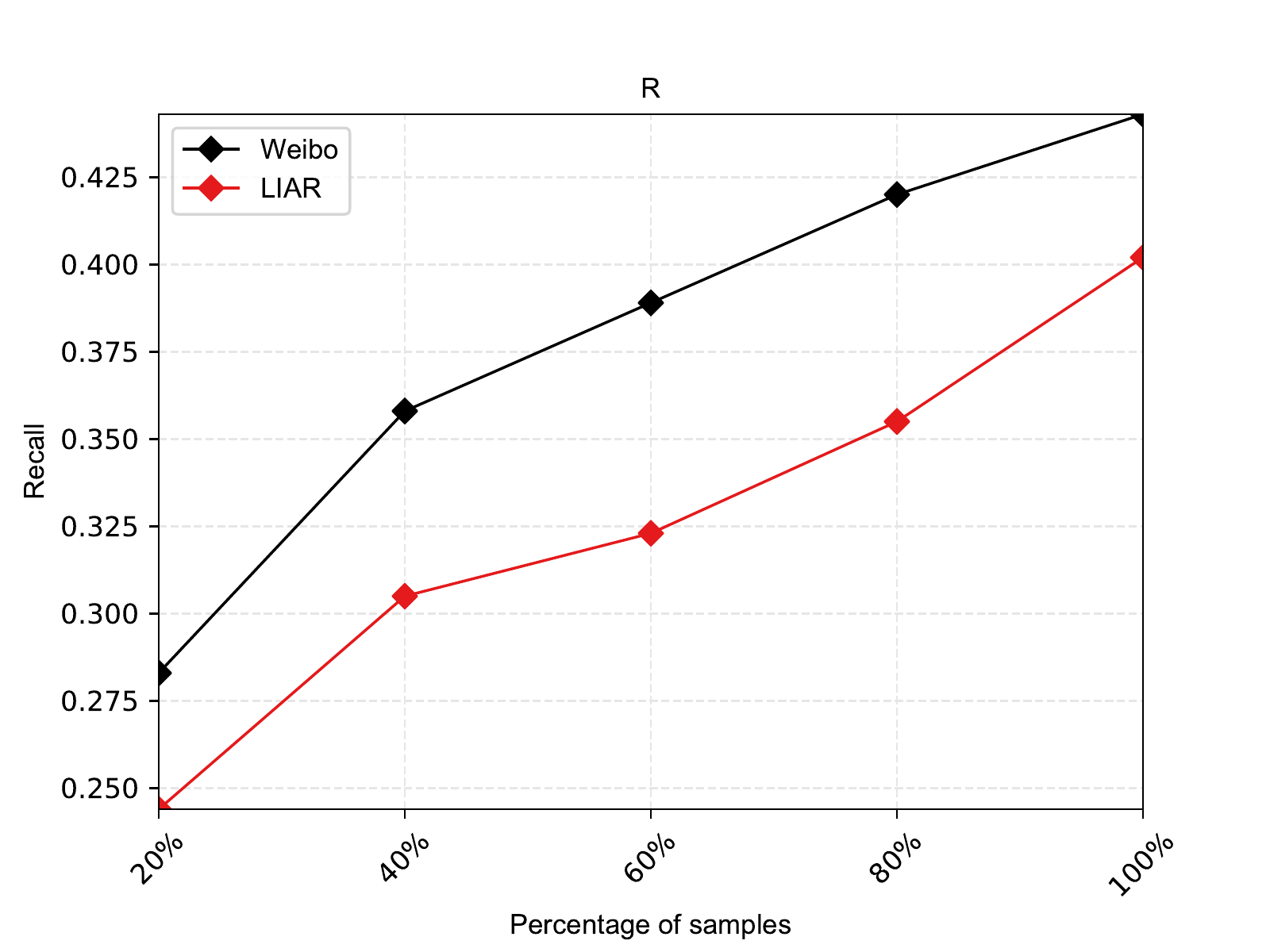}
    \end{minipage}
}
% 子图4： 图5(d)
\subfigure[F1-score]{
    \begin{minipage}[b]{0.23\textwidth}
    \includegraphics[width=1\textwidth]{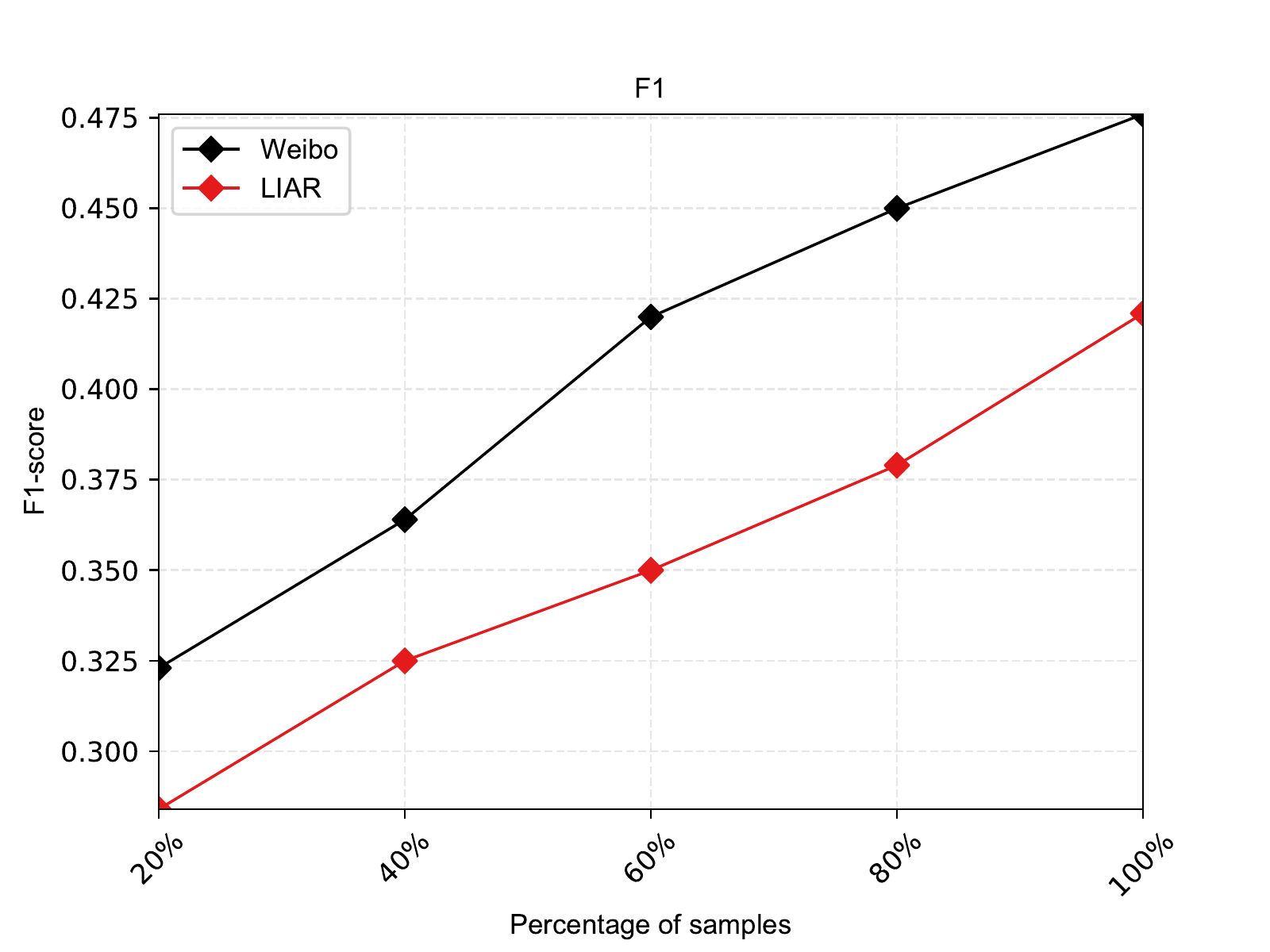}
    \end{minipage}
}
% 总标题
\caption{Comparison of the model performance under different partition of the amount of training samples} \label{fig:5}
\end{figure*}
\subsection{Performance of Components in ANSP} % Model Ablation
\label{model-ablation}
ANSP consists of multiple detachable components. To study the performance of different components in our model, we break down the model into several parts. The foundation in our model is the shared-private model without adversarial networks and the strategies, which is named as \textbf{basic}. \textbf{basic+Adv} model based on \textbf{basic} adds adversarial networks guided by reinforcement learning to Task 1. Similarly, on this basis, \textbf{basic +Adv+Ind} model adds additional \textbf{Ind} component to Task 2 as the independence strategy. \textbf{basic +Adv+Ind+Diff} model again adds \textbf{Diff} component as the differentiation strategy, where the \textbf{Ind} is the orthogonality constraints of common features and private features, and the \textbf{Diff} is used for reducing the similarity of common features and private features by negative KL-divergence.

We use the above models to test the validity of different components of our model. From Table \ref{tab:3}, we notice that the performance of the methods improves as the components increase. It indicates that the additional components in the basic method are helpful for credibility evaluation. Especially, the model with \textbf{Diff} outperforms obviously the model without \textbf{Diff}, which demonstrates the effectiveness of negative KL-divergence strategy in our model.

Furthermore, in order to evaluate the effectiveness of reinforcement learning in common feature extractor, we design two methods, i.e., the method of using adversarial networks with reinforcement learning and the method without reinforcement learning, to demonstrate its performance. The results are shown in Table \ref{tab:4}. We observe that the method of using adversarial networks with reinforcement learning achieves better performance, outperforming the latter with 3.7\% on accuracy, which indicates that reinforcement learning contributes to capturing common features from true and false information.

% 图6
\begin{figure*}
  \centering
  \subfigure[Experiments on LIAR]{
    \begin{minipage}[b]{0.4\textwidth}
      \includegraphics[width=1\textwidth]{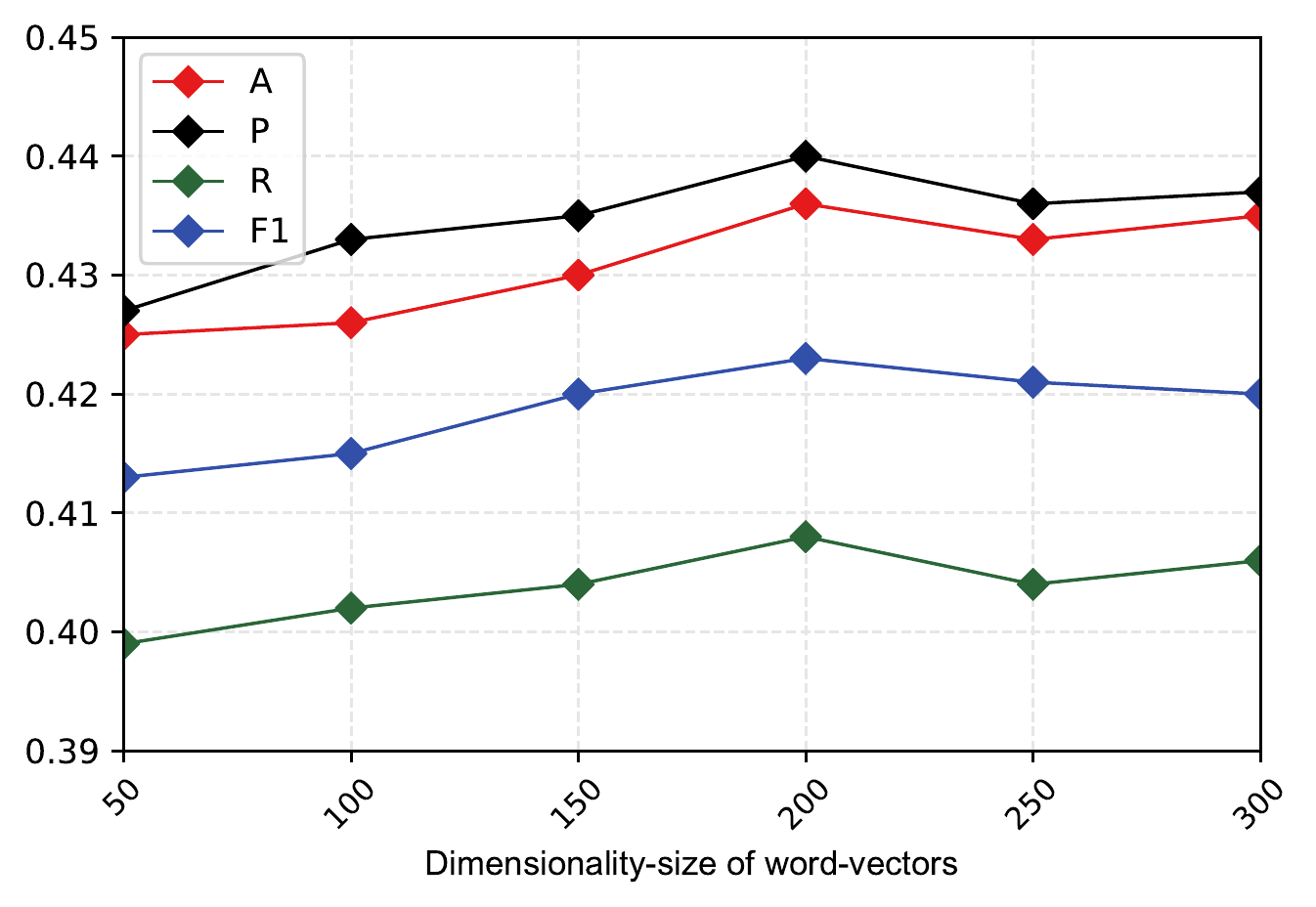}
    \end{minipage}
  }
  \subfigure[Experiments on Weibo]{
    \begin{minipage}[b]{0.4\textwidth}
      \includegraphics[width=1\textwidth]{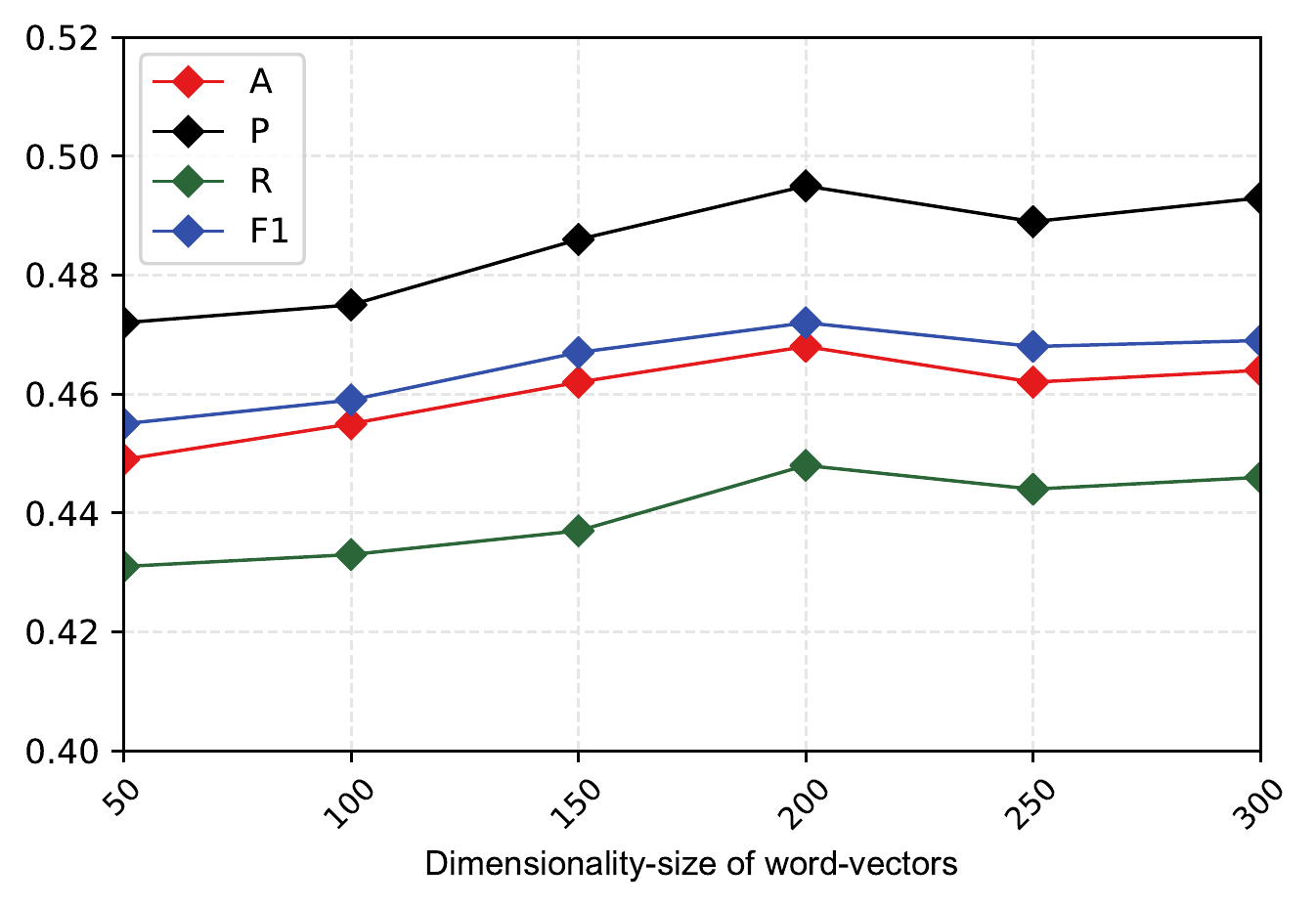}
    \end{minipage}
  }
  \caption{Comparison of the model performance under different dimensionality-size of word-vectors on the two datasets}\label{fig:6}
\end{figure*}

% 4.5
\subsection{Scalability Analysis of ANSP}

Besides analyzing the performance of ANSP, we study the scalability of the model in terms of varying amount of training samples, different dimensionality-size of word-vectors, and the convergence.

\subsubsection{Varying Amount of Training Samples}
We investigate the change of performance about ANSP with the varying amount of training samples based on evaluation metrics: accuracy, precision, recall, and F1-score. We first conduct experiments on 20\% of the training sample set and then evaluate the performance of the model with gradually adding the same number of training samples. As shown in Figure \ref{fig:5}, we observe that our model achieves the lowest performance on the earlier 20\% of the training sample set, which indicates that our model is inadequate training with a small training sample set since the model has a large number of parameters. Furthermore, we perceive that the performance is significantly improved as training samples increase, which implies that the model possesses well expansibility and enhancement. Additionally, as a result of the limitation to the number of training samples, the performance has not reached a stable maximum in whole experiments. In particular, around the point 100\% on the x-axis, the performance improvement is slower on Weibo than on LIAR. We predict the performance of our model on Weibo will soon touch the optimal value if we gave extra training samples, while the performance on LIAR will obtain continual improvements. The reasons may be that LIAR is relatively small, or our model can learn more valuable information from English dataset.

\subsubsection{The Different Dimensionality-size of Word-vectors}

We perform incremental testing of dimensionality-size of word-vectors from 50 to 300 on the LIAR and Weibo datasets. The results are shown in Figure \ref{fig:6} and we make the following observations:

\begin{itemize}
    \item According to four performance evaluation metrics, it is found that the effect of dimensionality-size of word-vectors on classification results is relatively small. Though the change of the performance on LIAR is more significant than on Weibo, and the floating range of the performance maintains tight 2\%. Specifically, the experimental results on accuracy are all near 43\% on LIAR and 45\% on Weibo in spite of different dimensionality-size.
    \item Even though the performance of the model does not change significantly under different dimensions, the model acquires the best performance in four metrics when the dimensionality-size is 200. We consider that the word embeddings include much more relevant semantics and less irrelevant semantics on this dimensionality-size. This is why we assign the dimensionality-size to be 200 in the final experiments.
\end{itemize}

\subsubsection{Convergence on the Training Set}
\label{convergence}
\begin{figure*}
  \centering
  \subfigure[The convergence of models on LIAR]{
    \begin{minipage}[b]{0.46\textwidth}
      \includegraphics[width=1\textwidth]{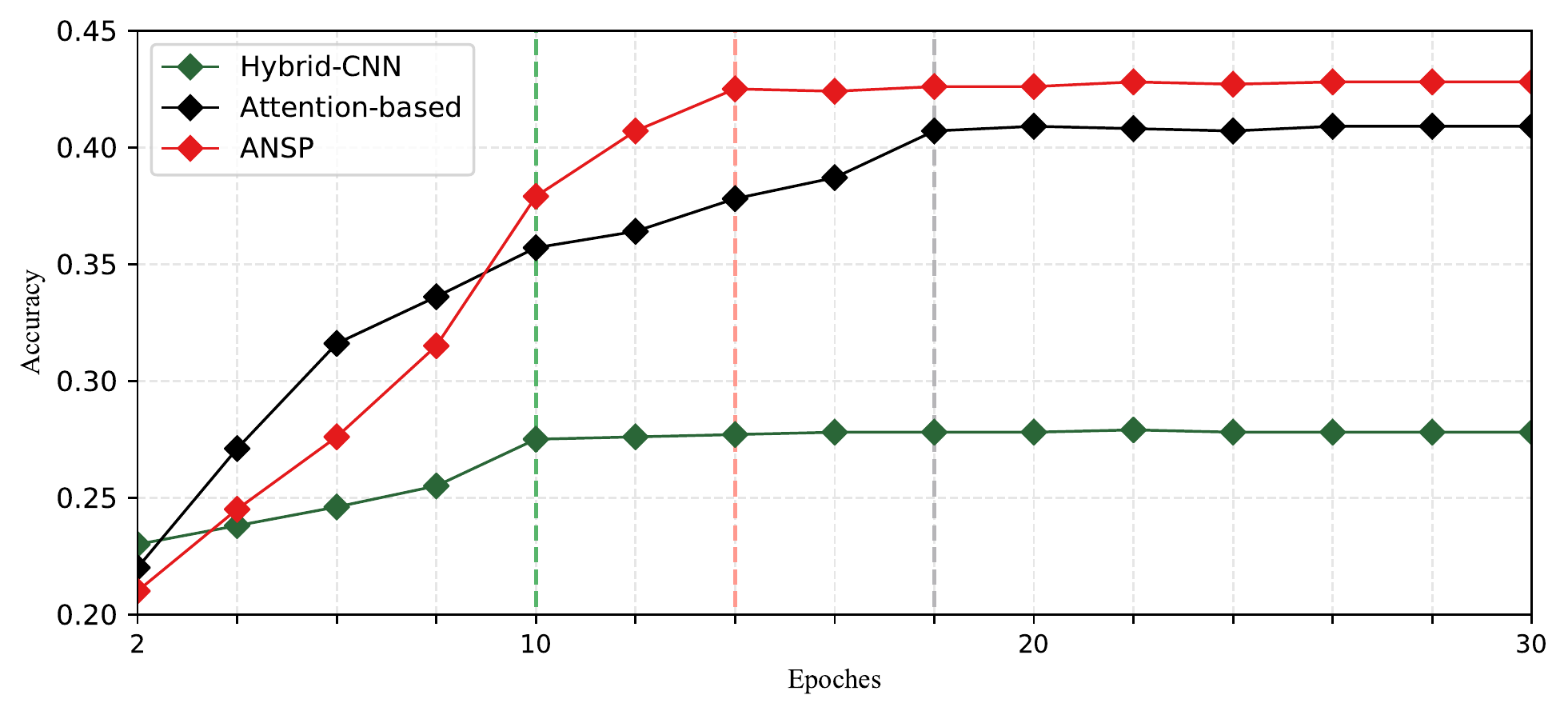}
    \end{minipage}
  }
  \subfigure[The convergence of models on Weibo]{
    \begin{minipage}[b]{0.46\textwidth}
      \includegraphics[width=1\textwidth]{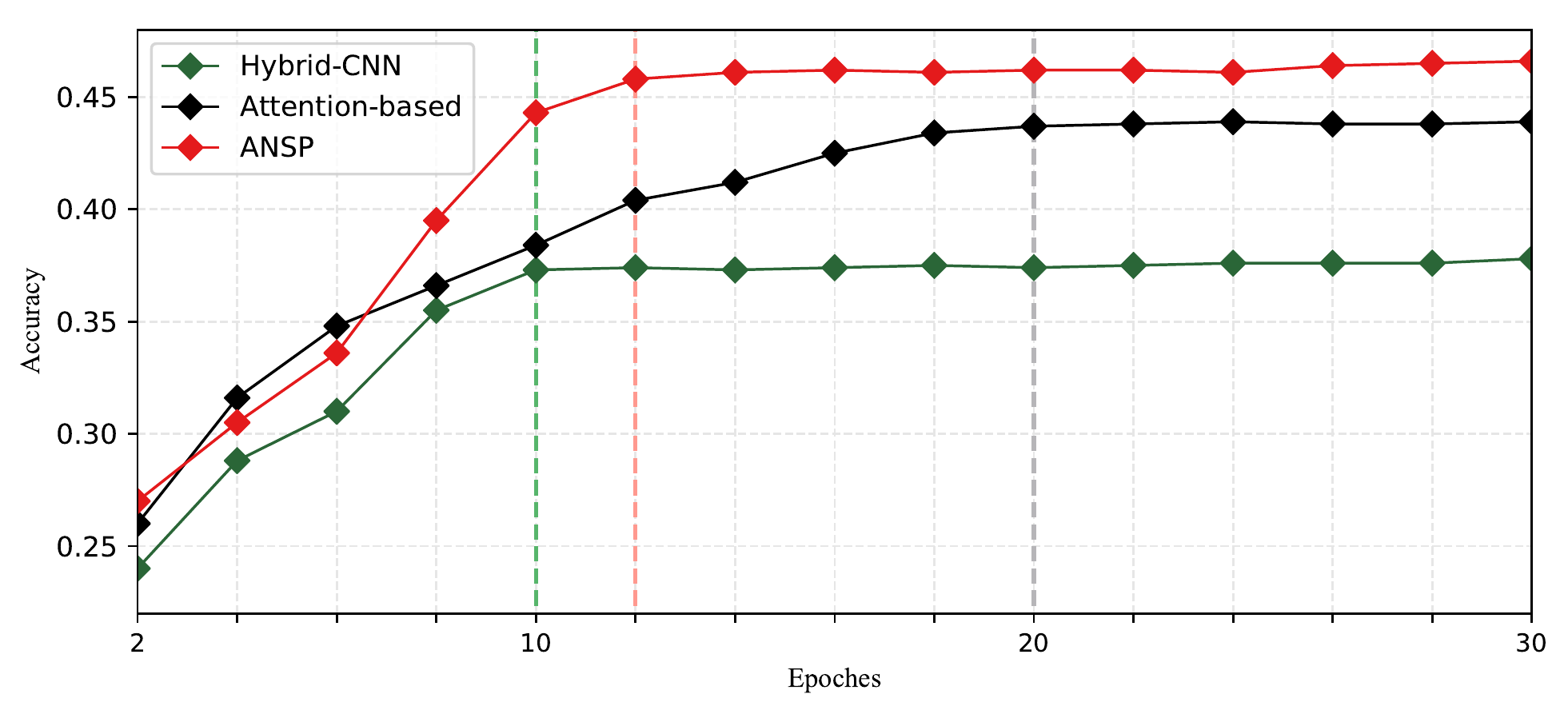}
    \end{minipage}
  }
  \caption{Convergence on the training set}\label{Fig:9}
\end{figure*}

Figure \ref{Fig:9} presents the training process of our model on LIAR and Weibo compared with several baseline algorithms. We observe that Hybrid-CNN method is faster in convergence speed (i.e., needing 8 epochs to reach stability on the two datasets) but poor performance, while Attention-based method has a lower convergence (i.e., needing 18 epoches on LIAR and 20 epoches on Weibo) speed than ANSP (i.e., needing 14 and 12 epoches on the two datasets, respectively). Therefore, we can conclude that our model not only achieves the best performance but also costs less time for training. Additionally, the convergence process of ANSP can be described as: the model converges slowly in initial training due to the continuous trial and error of reinforcement learning. After 6 epochs, the convergence of the model becomes faster and then reaches a stable state. At this time, our model learns stable and more distinguished features for credibility evaluation.
% 图6

\subsection{Robustness Analysis of ANSP on Twitter16}
In order to further testify the robustness of the model, we again conduct experiments on four-label Twitter16 dataset to evaluate its performance compared with the following baselines:

\textbf{SVM-TS:} A linear SVM model \cite{ma2015detect} utilizes the temporal characteristics to integrate various social context features for rumor identification.

\textbf{DTR:} A Decision-Tree-based Ranking method \cite{zhao2015enquiring} searches related posts based on inquiry phrases and clusters similar posts together, then rank these clusters containing disputed factual claims for rumor detection.

\textbf{RFC:} The Random Forest Classifier \cite{kwon2017rumor} employs user-based, structure-based, linguistics-based, and temporal features to track the precise changes in predictive powers across rumor features.

\textbf{GRU:} The RNN-based rumor detection model with gated recurrent unit \cite{ma2016detecting} are proposed by Ma et al. for capturing the variation of contextual information of relevant posts over time.

\textbf{BOW:} Ma et al. \cite{ma2017detect} set up BOW baseline method through using bag-of-words to represent semantic features of text and building a linear SVM for rumor detection.

\textbf{PTK:} A Propagation Tree Kernel model \cite{ma2017detect} captures high-order patterns to detect rumors through comparing the similarities between their propagation tree structures.

\textbf{cPTK:} On the basis of PTK, Ma et al. \cite{ma2017detect} successively develop Context-sensitive PTK by considering the propagation paths from the root of the tree to roots of subtrees to obtain the clues outside the subtrees for rumor detection.

The experimental results are shown in Table \ref{tab:twitter16} and we can obtain the following observations:

\begin{itemize}
  \item In all baseline methods, propagation-based cPTK and PTK gain the optimal performance, which exploit propagation trees based on kernel learning to encode the spread of a source tweet with complex structured patterns and flat information regarding content, user profiles, and time. But GRU capturing the textual and temporal features to represent tweet for rumor detection does not achieve better performance. These indicate methods fully integrated meta-data features are more effective than the methods without meta-data features.
  \item SVM-TS and RFC methods are both based on temporal traits to integrate hand-crafted features and acquire comparable performance. In spite of both methods learn deeply an extensive set of features, their performance is lower than ANSP largely owing to not all learned features are helpful to boost the performance of rumor detection.
  \item Our model outperforms all baseline methods. Specifically, our model gains 1.8\% improvements in accuracy compared with cPTK, which is because that our model based on multi-task learning not only captures deep credibility features but also more significantly, discovers differential features from these credibility features. Additionally, our model wins more obvious performance boosts on false rumor and true rumor in F1 score (i.e., 2.8\% and 2.5\%, respectively) than on non-rumor and unverified rumor, which results from the fact that our model separates common features between true and false rumors and discovered differential features are more conducive to detect true and false rumors instead of non-rumor and unverified rumor. According to the above-detailed analysis, we conclude that ANSP has favourable robustness and adaptability on different scenarios.
\end{itemize}

% 表  在Twitter16的性能
\begin{table}
\center
\scriptsize
  \caption{The performance evaluation of ANSP on Twitter16}
  \label{tab:twitter16}
  \setlength{\tabcolsep}{3.3mm}{
  \begin{tabular}{r|c|cccc}
    \toprule
    \multirow{2}*{Method} & \multirow{2}*{\textbf{Accuracy}} & \textbf{NR} & \textbf{FR} & \textbf{TR} & \textbf{UR}\\
    \cline{3-6}
    & & \textbf{F1} & \textbf{F1} & \textbf{F1} & \textbf{F1}\\
    \midrule
    DTR & 0.414 & 0.394 & 0.273 & 0.630 & 0.344\\
    SVM-TS & 0.574 & 0.755 & 0.420 & 0.571 & 0.526\\
    RFC & 0.585 & 0.752 & 0.415 & 0.547 & 0.563\\
    GRU & 0.633 & 0.772 & 0.489 & 0.686 & 0.593\\
    BOW & 0.585 & 0.553 & 0.556 & 0.655 & 0.578\\
    PTK & 0.722 & 0.784 & 0.690 & 0.786 & 0.644\\
    cPTK & 0.732 & 0.740 & 0.709 & 0.836 & 0.686\\
    \hline
    ANSP & \textbf{0.750} & \textbf{0.787} & \textbf{0.737} & \textbf{0.861} & 0.625\\
  \bottomrule
\end{tabular}
}
\end{table}

\subsection{Limitations}
\label{subsec:4.7}

From Table \ref{tab:twitter16} and Figure \ref{fig:diffr-types}, we can perceive that ANSP obtains different performance on different types of information. Specifically, the performance on types ``true" and ``false" is more remarkable than types ``half-true", ``half-false", etc. The reason is that our model discovers differential features through separating common features between true and false information. But this is coarse-grained due to ignoring the differential features between half-true and half-true types, which is a limitation to our model. In the future, to address this problem, we will focus on capturing the differential features among multiple types to improve the performance of different credibility types of information.

Additionally, although our model achieves noticeable performance on Twitter16 which obtain 0.75 and about 0.7 in accuracy and F1 score, respectively, our model only wins below 0.5 in accuracy and F1-score on LIAR and Weibo. We analyze the model and the datasets for the following reasons:

\begin{itemize}
    \item LIAR and Weibo are fine-grained and based on six-types. There are similar credibility features between adjacent types of information. For instance, ``mostly true" and ``half true" have many common features in the true part and the two types have slight differences in the false part. Since common feature extractor in our model aims at capturing common features between the binary of true and false information, ignoring subtle differences among these half-true and half-false types, which is the principal consideration for the model with inaccessibility satisfaction effect.
    \item The two datasets without comments and retweets lack users' attitudes to original tweets and the behavioral information to supplement the semantics of original tweets. These bring our model cannot obtain users' stances and sufficient context information for evaluation.
    \item The two datasets are relatively small. Specifically, though LIAR and Weibo include 12K and 18K tweets, respectively, the lengths of tweets in the datasets are below 90 and the bulk of tweets are short text. Simultaneously, the two datasets take below 3M and 10M disk space, respectively. Such small datasets with insufficient semantics may lead to over-fitting for most models.
\end{itemize}

\section{Conclusion}
\label{sec:5}
In this work, we proposed ANSP based on adversarial networks and multi-task learning to capture differential credibility features for information credibility evaluation. Specifically, adversarial networks guided by reinforcement learning are applied to capture common features, and two strategies, i.e., orthogonality constraints and KL-divergence, are used to reduce the common features from raw private features. Experiments first on two six-label datasets, i.e. LIAR and Weibo systematically demonstrated the effectiveness of ANSP and then on four-label Twitter16 testified robustness of the model. Finally, three limitations of ANSP are illustrated.

There are several directions for extending this work in the future. As we discussed in subsection \ref{subsec:4.7}, the first thing to be solved is how to effectively capture and represent slight differences among multiple credibility types of information, especially, half-true and half-false types. Secondly, for a large number of users to participate in the propagation of fake news, how to capture deeper user behavior features and integrate them into our model based on neural networks. Finally, to reduce labor costs of manual annotation, the problem of unlabeled samples classification by using shared spaces of shared-private models are worth exploring.

\section*{Acknowledges}{This work has received funding from ``the World-Class Universities(Disciplines) and the Characteristic Development Guidance Funds for the Central Universities"(PY3A022), Shenzhen Science and Technology Project(JCYJ201803061708\\36595), the National Natural Science Fund of China (No.F020807), Ministry of Education Fund Project ``Cloud Number Integration Science and Education Innovation" (No.2017B00030), Basic Scientific Research Operating Expenses of Central Universities (No.ZDYF2017006). We would like to thanks them for providing support.}

\section*{Acknowledgements}
The research work is supported by ``the World-Class Universities(Disciplines) and the Characteristic Development Guidance Funds for the Central Universities"(PY3A022), Shenzhen Science and Technology Project(JCYJ20180306170836595), the National Natural Science Fund of China (No.F020807), Ministry of Education Fund Project ``Cloud Number Integration Science and Education Innovation" (No.2017B00030), Basic Scientific Research Operating Expenses of Central Universities (No.ZDYF2017006).

\bibliography{example_paper}
\bibliographystyle{icml2019}

\end{document}